\documentclass[english,12pt,letterpaper]{article}
\usepackage[margin=1.0in]{geometry}
\usepackage{setspace}
\usepackage{amsmath} 
\usepackage{graphicx}
\usepackage{color}
\definecolor{Red}{rgb}{1,0.00,0.00}
\usepackage[round,numbers,sort&compress]{natbib} 
\usepackage{graphicx}

\usepackage{amssymb}

\usepackage{calc}
\usepackage{amsmath} 
\usepackage{extarrows} 

\usepackage[]{fontenc}
\usepackage[latin1]{inputenc}
\usepackage{epstopdf}
\usepackage[all]{xy}

\usepackage{float} 
\usepackage[]{caption} 
\newfloat{scheme}{thp}{lop}
\floatname{scheme}{Scheme}

\newfloat{sfigure}{thp}{lop}
\floatname{sfigure}{Supplemental Figure}

\usepackage{bm}
\newcommand{\FIGS}{./}
\newcommand{\pd}[2] {\frac{\partial #1}{\partial #2}}
\newcommand{\lpd}[2] {\partial #1/ \partial #2}

\newcommand*{\bs}{\backslash}
\renewcommand*{\to}{\rightarrow}
\newcommand*{\from}{\leftarrow}

\newcommand{\lfrac}[2] {#1 / #2}

\renewcommand*{\sec}{$\mathrm{s}^{-1}$}
 
\newcommand*{\Cm}{$\mathrm{Ca}^{2+}$} 
\newcommand*{\Cl}{$\mathrm{Cl}_{2}$} 
\newcommand*{\Mm}{$\mathrm{Mg}^{2+}$} 

\newcommand*{\Deg}{$^{\circ}$}
\newcommand*{\Celcius}{C}
\newcommand*{\uM}{$\mu\mathrm{M}$}
\newcommand*{\uL}{$\mu\mathrm{L}$}

\newcommand{\set}[1]{ \left\{ #1 \right\} }

\newcommand{\rate}[1]{\stackrel{#1}{\rightarrow}}
\newcommand{\revrate}[1]{\stackrel{#1}{\leftarrow}}

\newcommand{\rates}[2]{%
	\mathop{%
		\mathrel{%
			\raisebox{.4ex}{%
				\raisebox{-.8ex}{\hbox{\hspace{0.8ex}$\longleftarrow$}}
			 	\hbox{\hspace{-4.5ex}{$\longrightarrow$} }
			}%
		}%
	}^{#1}_{#2}
}

\title{~\\~\\~\\The cardiac Ca$^{2+}$-sensitive regulatory switch, a system in dynamic equilibrium}

\author{John~M.~Robinson\thanks{
	Corresponding author. Address:
	Department of Biochemistry and Molecular Genetics,
	University of Alabama at Birmingham,
	McCallum BHSB, Room 490,
	1918 University Blvd.,
	Birmingham, AL 35294
	Tel.:~(205)934-4004, Fax:~(205)975-4621; eMail:~jmr@uab.edu.} \\
Department of Biochemistry and Molecular Genetics, \\
Center for Computational Biology, \\
University of Alabama at Birmingham, Birmingham, AL\\
\and Herbert C. Cheung\\
Department of Biochemistry and Molecular Genetics, \\
University of Alabama at Birmingham, Birmingham, AL\\
\and Wenji Dong \\
School of Chemical Engineering and Bioengineering,\\
Department of Veterinary and Comparative Anatomy, \\
Pharmacology and Physiology, \\
Washington State University, Pullman, WA
}

\date{}
\begin{document}
\maketitle
\clearpage
\abstract{
The \Cm-sensitive regulatory switch of cardiac muscle is a paradigmatic example of protein assemblies that communicate ligand binding through allosteric change. The switch is a dimeric complex of troponin C (TnC), an allosteric sensor for \Cm, and troponin I (TnI), an allosteric reporter.  Time-resolved equilibrium FRET measurements suggest that the switch activates in two steps: a TnI-independent \Cm-priming step followed by TnI-dependent opening. To resolve the mechanistic role of TnI in activation we performed stopped-flow FRET measurements of activation following rapid addition of a lacking component (\Cm~or TnI) and deactivation following rapid chelation of \Cm. The time-resolved measurements, stopped-flow measurements, and \Cm-titration measurements were globally analyzed in terms of a new quantitative dynamic model of TnC-TnI allostery. The analysis provided a mesoscopic parameterization of distance changes, free energy changes, and transition rates among the accessible coarse-grained states of the system. The results reveal (i) the \Cm-induced priming step, which precedes opening, is the rate limiting step in activation, 
(ii) closing is the rate limiting step in deactivation, (iii) TnI induces opening,  (iv) an incompletely deactivated population when regulatory \Cm~is not bound, which generates an accessory pathway of activation, and (v) incomplete activation by \Cm---when regulatory \Cm~is bound, a 3:2 mixture of dynamically inter-converting open (active) and primed-closed (partially active) conformers is observed (15 \Celcius). Temperature-dependent stopped-flow FRET experiments provide a near complete thermo-kinetic parametrization of opening: the enthalpy change ($\Delta H = -33.4$ kJ/mol), entropy change ($\Delta S = -0.110$ kJ/mol/K), heat capacity change ($\Delta C_p = -7.6$ kJ/mol/K), the enthalpy of activation ($\delta^\ddagger = 10.6$ kJ/mol) and the effective barrier crossing attempt frequency ($\nu_{\mathrm{adj}} = 1.8 \times 10^{4}$~s$^{-1}$).
}

\clearpage

\section{Introduction}
The periodic contraction and relaxation of the heart is regulated by cytosolic \Cm~through an assembly of proteins consisting, minimally, of troponin C (TnC), troponin I (TnI), troponin T, tropomyosin, filamentous actin, and the head of myosin---an actin-binding ATPase~\cite{Robinson:2004dg, Gordon:2000kx, Kobayashi:2005fk}. The TnC-TnI assembly functions as a \Cm-sensitive regulatory switch as part of the larger regulatory assembly. \Cm~binding to the single functional regulatory site (site II) of the \Cm-receptor, TnC, is transduced as a delayed stochastic change in the isomerization state of troponin I (TnI).  The TnI isomerization event relieves contractile inhibition through the release of the inhibitory region of TnI (TnI-I, residues 130-149) from actin~\cite{Tao:1990fk}. Inhibition is due the regulated association TnI-I with actin that places tropomyosin in a blocking position, which inhibits inorganic phosphate (P$_i$) release from myosin. P$_i$ release is required to form the strongly bound, force generating, acto-myosin complex~\cite{Pate:1998fk}.  Activation of the TnC-TnI assembly involves an inter-helical rearrangement, called opening, in the \Cm-binding EF-hand (helix-loop-helix) motifs in the N-domain of TnC. The pivoting of helices B and C away from the central helix (D)~\cite{Herzberg:1986fv} enables previously buried hydrophobic residues in the B, C and D helices to associate with hydrophobic residues in the regulatory helix of TnI (TnI-R, residues 150-165) that is contiguous with TnI-I. In contrast to the fast skeletal muscle TnC, which has two functional regulatory \Cm~binding sites (sites I and II), in the cardiac system, TnI-R is required for the opening of TnC~\cite{Spyracopoulos:1997nt, Dong:1999mt}.

The  cardiac TnC-TnI assembly is an allosteric system. Classical models of allostery, such as the  Monod-Wyman-Changeux~\cite{Wyman48} and Koshland-Nemethy-Filmer~\cite{Koshland:1966kx} models, and generalized models using linked functions~\cite{Wyman:1972a} or conditional free energies~\cite{Weber:1972lr}, consider systems that are in equilibrium. Allostery is the long-range coupling between distinct regions of a macromolecule.  The classic requirements that an allosteric system be oligomeric and symmetric have given way to a new definition of allosteric systems---systems where the binding of one ligand affects the affinity of a second ligand. This definition includes monomeric multidomain proteins, where the second ligand is second protein domain, and macromolecular assemblies, where the second ligand is another member of the assembly. The cardiac TnC-TnI assembly belongs to this last class.

Over the past decade, the role of dynamics in allosteric regulation has drawn increasing interest~\cite{Kern:2003uk}. The protein is now seen as a fluctuating entity that dynamically exchanges among a large number of microstates~\cite{Kern:2003uk, Hilser:2006pb}. These microstates can be organized into regions of local stability, called conformational substates~\cite{Matoba:2003fs} or macrostates, that correlate with functional activity. Dynamics within a macrostate involves fluctuating motions of individual atoms, residues, and groups of residues on the picosecond to microsecond timescale.  These fluctuations are the basis for the entropy of a macrostate~\cite{Prabhu:2003pd}. A second class of dynamics, which we call \emph{allosteric dynamics}, involves random jump-like transitions among macrostates on the microsecond to second timescale. The focus of this study is the allosteric dynamics of the cardiac TnC-TnI assembly.

F\"orster Resonance Energy Transfer (FRET) provides a powerful tool to study ligand binding-induced inter-domain distance changes (on the near \AA~scale) in protein assemblies in their native environment~\cite{Clegg:2006a, Lakowicz:2006a}. In contrast to conventional fluorescent assays that report change in the local environment of the probe, FRET provides a clear metric for allosteric change---a change in the mean inter-probe distance. Time resolved FRET measurements can quantitate the \emph{distribution} of inter-probe distances~\cite{Cheung:1991lr}.  Used in experiments involving dynamic change through some perturbation of the system, FRET provides what has been termed \emph{structural kinetics}~\cite{Dong:2008lf} with structural monitoring of \Cm~regulated allosteric dynamics of the TnC-TnI assembly.

An unfortunate property of measured perturbation-induced relaxation rates is that they are \emph{functions} of the elementary forward and backward transition rate constants that govern the allosteric dynamics of a system~\cite{Bernasconi:1976a}. One approach to resolving the component elementary rate constants from observed relaxation rates is to perform a set of kinetic experiments that contain overlapping information. When two or more experiments jointly depend on underlying elementary transition rate parameters, a combined (or global) analysis of the experiments may resolve the rate parameters~\cite{Beechem:1992fk, Robinson:2003hk}. This resolution is usually not possible when experiments are analyzed independently and empirically.

Here, we have pursued such a global strategy by using a previously characterized FRET reporter system on TnC~\cite{Dong:1999mt} in a set of experiments to parametrize the structural kinetics of the cardiac TnC-TnI assembly. Global analysis of the experiments required that we formulate a mesoscopic model of TnC-TnI allosteric dynamics during the activation and deactivation stages of the signaling cycle. We performed time resolved FRET measurements of Mg$^{2+}$- or \Cm-saturated~(apo/holo) and TnI bound/unbound samples; stopped-flow FRET measurements of \Cm-induced activation, TnI-induced activation, and \Cm-chelation induced deactivation; and a FRET-monitored \Cm-titration measurement. The measurements were analyzed in terms of the allosteric model to provide distance changes, free energy changes, and the elementary rate constants for transitions among the accessible coarse-grained states of the system. The analysis was used to differentiate between two mechanistic models of TnC-TnI activation/deactivation. The experiments were used to construct the basin-limited portions of the free energy landscape that supports signaling. To obtain a more complete thermo-kinetic parameterization of the TnC opening transition, the stopped-flow FRET measurements were repeated for a set of temperatures. The recovered transition rates for opening/closing were subjected to van't Hoff and modified Arrhenius analyses that quantitated the heat capacity, enthalpy change and entropy change of opening, the forward and backward enthalpy of activation, and the forward and backward effective barrier crossing attempt frequencies. Our measurements suggest that the cardiac TnC-TnI assembly is in dynamic equilibrium among its macrostates both when regulatory \Cm~is bound and unbound. When \Cm~is bound, the probability that the TnC-TnI complex is active is only 60\%.

\section{Methods}

\subsection{Sample preparation}
The construction, purification methods, and validation of  TnC(F12W/N51C/C35S/C84S) (abbreviated name, TnC(12W/51C)) as well as tryptophan-less cardiac troponin I, TnI(W192F) (abbreviated name, TnI(W-)), used in this study have been described~\cite{Dong:1999mt}. To remove unlabeled protein from the reaction mixture (\textit{i.e.} obtain 100\% labeling efficiency) the mixture was fractionated as described~\cite{Dong:2006lr}. The binary troponin complex was prepared by incubating TnC(12W/51C$\pm$AEDANS) (2 \uM) with an 2-fold excess of TnI(W-) (4 \uM) on ice for 20 minutes. 

\subsection{Spectroscopic measurements}
Except where noted, all measurements were performed at $15 \pm 0.1 $ \Celcius. Samples were prepared in standard buffer (SB): 50 mM 3-(N-mopholino)propanesulfonic acid (MOPS), pH 7.0, 1 mM dithiothreitol (DTT), 5 mM Mg\Cl, 0.2 M KCl; working buffer (WB): SB +  2 mM ethylene glycol-bis-($\beta$-aminoethyl ether)-N,N,N',N'-tetraacetic acid (EGTA); or activating buffer (AB): SB + 160 \uM~\Cm~(pCa, 3.8). A stable TnC-TnI complex is formed in WB with 2-fold molar excess TnI as shown previously~\cite{Dong:2000mm}, where, using an engineered TnI(W150), TnI-binding-induced changes in the Trp lifetime were saturated with a 1.2 molar excess of TnC in the absence of Mg$^{2+}$.  \Cm~titrations of the FRET distance were performed as described \cite{Robinson:2004dg} with the following modifications. A microtitrator (ISS, Champaign, IL) delivered 90 successive 5 \uL~injections of a \Cm-EGTA solution into reconstituted binary troponin complex ([TnC] =  1 \uM) in WB (1.0 mL starting volume). The FRET donor, tryptophan, was excited at 295 nm; the FRET-quenched donor fluorescence was monitored at 340 nm (monochromator slit width, 2 $\mu$m).

The ensemble distribution of the inter-probe distance was obtained from the time-resolved multi-exponential decay of the FRET donor Trp12 as described in detail \cite{Robinson:2004dg} with the following modifications. The time-resolved fluorescent decays of Trp12 were collected using an IBH 5000U single photon counting time domain fluorescent lifetime instrument. Excitation was with a  295 nm pulsed LED. Isolated TnI(W-) in WB provided background fluorescence and scattering, which was subtracted from donor-containing samples. (Failure to subtract background fluorescence, which is particularly problematic when using Trp as a FRET donor, produces reported FRET distances that are artificially low.)

Perturbation-induced time-dependent  FRET distances were obtained as described~\cite{Dong:2003pr} using a Kintek F2004 stopped flow mixing spectrometer (1.8-ms instrument dead time).  Deactivation kinetics: rapid mixing of a protein in AB with an equal volume of WB (post-mixing: 2 \uM~TnC, 1 mM EGTA). \Cm-induced activation: rapid mixing of the binary troponin complex in SB +  minimally buffered (30  \uM~EGTA was mixed with an equal volume of SB +  500 \uM~\Cm~(post-mixing: 2  \uM~TnC, 250 \uM~\Cm). TnI-induced activation: rapid mixing of a \Cm-saturated TnC in AB with an equal volume of TnI in AB. A reduced concentration of TnI (post-mixing: 0.5 \uM~TnC, 1 \uM~TnI) was used due to the low solubility of TnI in 0.2 M KCl. FRET distances were calculated from the average of eight to ten tracings of concentration matched donor-only and donor- acceptor samples. Mean FRET distances recovered from the time-resolved data provided distance calibration standards for the stopped-flow measurements. Mock injections, where the protein was mixed with the same buffer, were used for calibration and to exclude the possibility of dilution-induced artifacts in the measured fluorescence.

\subsection{Data analysis} 
The time-resolved FRET experiments were fit as described~\cite{Robinson:2004dg}. Using the extensible global analysis software, GlobalCurve \cite{Robinson:2002yh}, the stopped flow FRET data were first independently fit  to an empirical sum of exponentials then globally fit to the kinetically coupled models of activation and deactivation (Eqs.~\ref{eq:activation_a}, \ref{eq:activation_b}, and \ref{eq:deactivation}).  Global fitting required custom-written routines to solve the set of coupled ordinary differential equations for probability distribution over the system-state space (see Modeling section) during activation ($a$) and deactivation ($d$), 
\begin{equation}
\begin{array}{rcl}
\frac{d\mathbf{P}^{a}(t)}{dt}   &  = &  (\alpha_1 \mathbf{A}_{1} + \alpha_2 \mathbf{A}_{2}) \mathbf{P}^{a}(t) \\
 \frac{d\mathbf{P}^{d}(t)}{dt} &  = &  \mathbf{D} \mathbf{P}^{d}(t)
\end{array},
\label{eq:rate_eq}
\end{equation}
where $\alpha_i$ is the relative species flux through path $i$. $\mathbf{P} = \set{P_{\bs 00} , P_{\bs 10}, P_{\bs 11} }$ is the time-dependent marginalized probability distribution  for finding TnC in state $s_1$ and TnI in state $s_2$ (\emph{i.e.} $S' = (\bs s_1 s_2)$), obtained by summing over $s_0$
\begin{equation}
P_{\bs s_1 s_2} = P_{0 \bs s_1 s_2} + P_{1 \bs s_1 s_2}.
\label{eq:marginal}
\end{equation}
The distribution is normalized, $\| P_{s_0 \bs s_1 s_2} \|  = 1$. The kinetic transition matrix for the first path of activation is
\begin{equation}
\mathbf{A}_1 =
\left( \begin{array}{ccc} -k_{1} & k_{-1} & 0 \\ 
k_{1} & -(k_{-1} + k_{2}) & k_{-2} \\
0 & k_{2}  & -k_{-2}
\end{array} \right).
\label{eq:activation_matrix}
\end{equation}
The kinetic transition matrix for deactivation and the second path of activation (with $k'_{-3}$ = 0.1 \sec~substituted for $k_{-3}$) is
\begin{equation}
\mathbf{D} = \mathbf{A}_2 =
\left( \begin{array}{ccc} -k_{3} & k_{-3} & 0 \\ 
k_{3} & -(k_{-3} + k_{2}) & k_{-2} \\
0 & k_{2}  & -k_{-2}
\end{array} \right).
\label{eq:deactivation_matrix}
\end{equation}
The equations were numerically integrated (Gear method, IMSL) subject to the following initial conditions. The initial condition for the \Cm-induced activation kinetics is the equilibrated system under deactivating conditions, $P^{a}_{S'_i}(0) = P^{d}_{S'_i}(\infty) \equiv P^{\mathrm{eq},d}_{S'_i} \in \mathcal{N}(\mathbf{D})$, where $\mathcal{N}(\mathbf{D})$ is the null space of $\mathbf{D}$, obtained from the singular value decomposition of $\mathbf{D}$, and $P^{\mathrm{eq},d}_{S'_i}$ is the equilibrium distribution of the protein state $S'_i$ under deactivating [\Cm]. The initial condition for TnI-induced activation is obtained by  setting $k_{2} = 0$ in $\mathbf{A}_1$, giving $\set{P^{a}_{S'_i}(0)} = \set{  (1+\frac{k_{1}}{k_{-1}} )^{-1} ,\frac{k_{1}}{k_{-1}} (1+\frac{k_{1}}{k_{-1}} )^{-1} , 0 }$.  The initial condition for the \Cm-chelation-induced deactivation kinetics is the equilibrium distribution under saturating [\Cm], $P^{d}_{S'_i}(0) = P^{a}_{S'_i}(\infty) \equiv P^{\mathrm{eq},a}_{S'_i} \in \mathcal{N}(\mathbf{A})$.  To deal with calibration error caused by noise in the stopped flow measurements and experimental error in the recovered mean FRET distance from the time-resolved measurements, small adjustable distance correction factors were added to the models during global fitting.

The observed FRET distance in the stopped flow measurements and in the \Cm-titration measurement is the population weighted average 
\begin{equation}
\left< d \right>(\mu) = \sum_{\set{S'_i}} P_{S'_i}(\mu) d_{S'_i},
\label{eq:R}
\end{equation}
where $\set{d_{S'_i}} = \set{d_{\bs 00}, d_{\bs 10}, d_{\bs 11}}$ are the species-associated FRET distances. The equilibrium system-state distribution at the \Cm~chemical potential $\mu$ is given by the Boltzmann equation
\begin{equation}
P^{\mathrm{eq}}_{S_i}(\mu) = \frac{e^{-\beta G_{S_i}(\mu)}}{\sum_{\set{S_j}}e^{-\beta G_{S_j}(\mu)}},
\label{eq:Boltzmann}
\end{equation}
where $\beta = 1/R T$, $R$ is the Boltzmann gas constant, and $T$ is absolute temperature in K. The set of free energies of the accessible system-states $\set{G_{S_i}(\mu)}$ parametrize the basin-limited portion of the free energy landscape.

\section{Modeling}

We have developed a mesoscipically-resolved theory of nonequilibrium allostery suitable for describing the isomerization dynamics of TnC and TnI as part of the cardiac TnC-TnI assembly. The theory provides linked models of the activation and deactivation components of the signaling cycle, and it can be used to combine both structural and kinetic information into an integrated analysis of \Cm-mediated signaling. To construct a model of TnC-TnI allostery assumptions must be made about the nature and number of allosteric transitions within the system. Many allosteric proteins, including TnC and TnI, are adequately described as bi-meta-stable~\cite{Monod:1965kl, Volkman:2001bs}, meaning they possess two macrostates: an active and an inactive isomer. The kinetics of \Cm-induced structural changes in TnC-TnI are well fit by a three-step sequential model involving two isomerization steps~\cite{Dong:1997lr}. The two isomerization steps can be interpreted as individual two-state isomerization events in the component proteins TnC and TnI.

Here, the allosteric protein is represented as a union of two domains---an input domain and an output domain. Each domain can be in one of two macrostates. Ligand binding to the input domain involves direct steric interaction, while the interaction between the input and output domains is allosteric. From the steric effects of molecular association~\cite{McCammon:1987a}, the state of the input domain is absolutely correlated with ligand binding. This assumed correlation satisfies the lock-and-key~\cite{Fischer_1894} and induced-fit~\cite{Koshland:1966kx} models. [The lock-and-key process is just an induced-fit process without structural change.] The system description is visualized using Scheme~\ref{scm:model}, which enumeraties all possible combinations of input domain, output domain, and the ligand binding status of the input domain.
Each domain is either inactive (square) or active (circle). The input domain is on the left. Bound ligand is represented as a filled circle. Of the eight possible states, the four that are allowed are shown with a grey background. For these, the active input state (circle) is ligand-bound and the inactive input state (square) is ligand-unbound---the state of input domain is absolutely correlated with ligand binding. In contrast, the output domain can be either active (circle) or inactive (square) both when ligand is bound or unbound---it is a ``pre-existing equilibrium" process~\cite{Weber:1972lr, Freire:1999fk}. Our model of allostery is thus a hybrid of pre-existing equilibrium and induced-fit models. The model describes allostery in all members of an assembly of allosteric proteins, provided that the output of a protein is viewed as the input ``ligand'' for the immediately downstream protein. In this way, inter-molecular signaling is treated as a steric interaction. For members of an allosteric assembly that do not function as the principal ligand receptor, only the output state of the protein needs to be specified because the state of its input is inferred from the state of the output of the upstream protein.

The TnC-TnI complex  is a \emph{thermodynamic system} of three components: one \Cm, one TnC, and one TnI. The configuration of the entire system, called the system-state, is represented as a binary string, $S = (s _0 \bs s _1 s _2)$, which is a labeling scheme comprised of bits that specify the state of each component. The zero'th bit $s_0$ gives the ligand binding status of the \Cm-regulatory site (loop II) of TnC (0 = \Cm-not bound; 1 = \Cm-bound);  the first bit specifies the output state of TnC, and the second (terminal) bit specifies the output state of TnI. For bits representing the output state of a protein (bits to the right of $\bs$), 0 denotes the inactive state and 1 denotes the active state.

At the mesoscopic level, allosteric signaling in the TnC-TnI assembly is a stochastic process because ligand-binding and protein isomerization events are random thermally-activated transitions.  Continuous-time random processes are governed by an evolution equation, called the master equation, for the probability $P_{S}(t)$ to find the system in state $S$ at time $t$~\cite{Schnakenberg:1976a}
\begin{equation}
\frac{dP_{S}(t)}{dt} = \sum_{\rho,S'}[W_{\rho}(S'|S)P_{S'}(t) -  W_{-\rho}(S'|S)P_{S}(t) ].
\label{eq:Master}
\end{equation}
The quantities $W_{\rho}(S'|S)$ and $W_{-\rho}(S'|S) = W_{\rho}(S|S')$ denote, respectively the rates of the transitions $S' \rate{\rho} S$ and $S' \revrate{-\rho} S$ for the elementary processes $\rho = 1,2,\ldots, r$, of which there are $r=8\cdot7/2 = 28$.  Many of the fundamental properties of the nonequilibrium behavior of the TnC-TnI assembly can be investigated and understood in terms of the Markov network  of the system---a graph associated with the master equation (Eq.~\ref{eq:Master}). In a Markov network, system-states $S$ are represented as vertices (nodes), and the elementary (bi-directional) transitions $\rho$ are represented as edges.

Four carefully considered assumptions, afford a significant reduction in the number of elementary transitions and the number of independent  kinetic/thermodynamic parameters needed to describe \Cm-induced signaling in the TnC-TnI assembly. \emph{(i)} Only one component can switch at a time. Thus, system-state changes $S' \to S$ involve transitions in a single bit $s_j$. Single bit transitions  are  represented by $\sigma_{j}^{+} \equiv (s_j: 0 \to 1)$, and $\sigma_{j}^{-} \equiv (s_j: 1 \to 0)$. The Markov network of theTnC-TnI assembly undergoing only single bit transitions is shown in Fig.~\ref{fig:network_models}a. The number of elementary transitions are reduced to $r = 8\cdot3/2= 12$. \emph{(ii)} The rate constants for each elementary transition $\rho$ obey detailed balance
\begin{equation}
W_{\rho}(S'|S)P^{\mathrm{eq}}_{S'} =  W_{-\rho}(S'|S)P^{\mathrm{eq}}_{S},
\label{eq:detailed_balance}
\end{equation}
where from Eq.~\ref{eq:Boltzmann},
\begin{equation}
\lfrac{P^{\mathrm{eq}}_{S}}{P^{\mathrm{eq}}_{S'}} = e^{-\beta (G_{S} - G_{S'})}.
\label{eq:probs}
\end{equation}
$P^{\mathrm{eq}}_S$ is the probability that the equilibrated system occupies the system-state $S$. We can identify a set of six minimal network circuits $\set{\mathcal{C}_1, \mathcal{C}_2, \ldots \mathcal{C}_6} $, where travel along allowed elementary transitions $\rho$ returns the system to its starting position.  Each minimal circuit contains four nodes and consists of transitions in two bits. The cyclic flows $C^{+}_l$ and $C^{-}_l$ on the circuit $\mathcal{C}_l$ involve, respectively, clockwise and counter-clockwise travel in the circuit. Conservation of free energy dictates that the net free energy change along each circuit $\mathcal{C}_l$ is zero. From conservation of free energy, Eq.~\ref{eq:detailed_balance}, and Eq.~\ref{eq:probs}, we find that for each circuit $\mathcal{C}_l$ the product of transition rates for $C^{+}_l$ must balance the product of transition rates for $C^{-}_l$,
\begin{equation}
\prod_{i \in C^+_l} W_{i}=\prod_{j \in C^-_l} W_{j}.
\label{eq:MDB}
\end{equation}
Transition rates are thus \emph{macroscopically detail balanced}. Circuits that involve transitions in all three bits can be defined, but they do not provide additional restraints on the transition rates.

Constraints \emph{(i)} and \emph{(ii)} apply to both simple systems, such as coupled chemical reactions, and more complex systems, such as assemblies of allosteric proteins. The TnC-TnI assembly, being a complex spatially extended system, is subject to two additional constraints. \emph{(iii)} Each component of the assembly knows only its local environment. This is a defining feature of complex systems~\cite{Bar-Yam:1992}. It implies, for example, that TnI has no direct knowledge of whether regulatory \Cm~is bound to TnC; rather, \Cm~binding is communicated to TnI by an allosteric change in TnC. Incomplete knowledge manifests as nearest-neighbor-limited influence, causing certain transition probabilities to be degenerate
\begin{equation}
W( \sigma_{j}^{\pm} | s_{0} \ldots s_{j-1} s_{j+1} \ldots s_{n} ) = W ( \sigma_{j}^{\pm} | s_{j-1} s_{j+1} ),
\label{eq:nn_influence}
\end{equation}
where $n=3$. Here, the rate of the transition involving bit $s_j$, $W(s_0 \bs s_1 \ldots s_{j-1} \sigma_j^{\pm} s_{j+1} \ldots s_n)$, has been rewritten to show explicitly how the rate depends on the remaining bits $s_{k \neq j}$: $W(\sigma_j^{\pm} | s_0 s_1 \ldots s_{j-1} s_{j+1} \ldots s_n)$. In the TnC-TnI complex, the nearest-neighbor assumption (Eq.~\ref{eq:nn_influence}) dictates that the transition rates for TnI switching do not directly depend on whether regulatory \Cm~is bound to TnC (i.e. $W( 0 \bs 1  \sigma_{2}^{\pm}) = W( 1 \bs 1  \sigma_{2}^{\pm})$) and that the transition probabilities for \Cm~binding/release do not directly depend on the isomerization state of TnI (i.e. $W ( \sigma_{0}^{\pm} \bs s_1 0) = W( \sigma_{0}^{\pm} \bs s_1 1)$). Degenerate transition rates evidently originate in the constrained topology of the free energy landscape of the system. The forward and reverse transition probabilities $W_{\rho}$, $W_{-\rho}$ along each edge $\rho$ depend on the geometry of the two-basin \emph{free energy surface} of that edge (Supplemental~Fig. 1) as well as the basin-associated friction coefficients~\cite{Hanggi_1990a, Berezhkovskii:2005lr}. The concept of a one-dimensional \emph{free energy surface} along the reaction coordinate of a transition is generalized to the composite \emph{free energy landscape}, consisting of the set of free energy surfaces over the reaction coordinates of all transitions $\rho$ in the Markov network. \emph{(iv)} The two system-states with inactive TnC but active TnI---$(0 \bs 01)$, $(1 \bs 01)$---are recognized as high energy states, where the TnI-R is inserted into an unexposed hydrophobic pocket on TnC. Because of their high energies, the probability that these states are visited is negligible. These states are deleted from the network, as are the transitions involving them. In summary, we have introduced four assumptions that reduce the number of accessible network states from eight to six and constrain the transition rates that govern system dynamics.

We assume that transition rates $\set{W_{\pm\rho}}$ are governed by the law of mass action and that the transition rates obey the empirical Arrhenius Law---an exponential dependence of rate on activation energy. Out of convenience we assume that the rate of \Cm~dissociation from partially activated TnC-TnI and fully-activated TnC-TnI is intermediate ($\sim 10/\mu\mathrm{s}$). Below, we relax this assumption and show that the assumption does not restrict the results. In Fig.~\ref{fig:network_models}b the parent Markov network (Fig.~\ref{fig:network_models}a) is divided into two sub-networks---networks for the activation and deactivation components of the signaling cycle. The transition rates $\set{W_{\pm\rho}}$ for the protein isomerizations are given in terms of the kinetic rate constants $k_{\pm i}, i = 1,2,3$.  The activation and deactivation sub-networks  are kinetically linked through common $k_2$, $k_{-2}$, $k_3$, and $k_{-3}$. System states $(0 \bs 01)$ and $(1 \bs 01)$ are eliminated and degenerate transition rates are shown---the transition rates of TnI activation ($(0 \bs 10) \to (0\bs11)$) and ($(1 \bs 10) \to (1\bs11)$) are a common $k_2$; the transition rates of TnI deactivation ($(0 \bs 10) \from (0\bs11)$) and ($(1 \bs 10) \from (1\bs11)$) are a common $k_{-2}$. $W_{\pm \rho}$ for the elementary \Cm~binding transitions are not considered. Instead, their quotients, the bi-molecular  association constants $K_0$, $K_1$, and $K_2$ (in M$^{-1}$) are defined (see Appendix B) for \Cm~binding/release, to the system-states  $(\bs 00)$, $(\bs 10)$, and $(\bs 11)$. For both activation and deactivation, two pathways, the anticipated dominant (red) and minor (blue) pathways, are identified. Transitions common to both pathways are shown in green. The minor pathway for activation (deactivation) involves the binding (release) of \Cm~from a partially activated (deactivated) TnC-TnI complex. The dominant and minor pathways of activation are given, respectively, by the kinetic schemes
\begin{subequations}
\begin{equation}
0 \bs 00 \xLeftrightarrow{ K_{0}[ \mathrm{Ca}^{2+}] }  1 \bs 00 \rates{k_1}{k_{-1}} 1 \bs 10 \rates{k_2}{k_{-2}} 1 \bs 11   
\label{eq:first_activation_a}
\end{equation}
\begin{equation}
0 \bs 00 \rates{k_3}{k_{-3}} 0 \bs 10 \xLeftrightarrow{ K_{1}[ \mathrm{Ca}^{2+}] } 1 \bs 10 \rates{k_2}{k_{-2}} 1 \bs 11   
\label{eq:first_activation_b}
\end{equation}
\label{eq:first_activation}
\end{subequations}
The dominant and minor pathways of deactivation are given, respectively, by
\begin{subequations}
\begin{equation}
0 \bs 00 \rates{k_3}{k_{-3}} 0 \bs 10 \rates{k_2}{k_{-2}}  0 \bs 11 \xLeftrightarrow{ K_{2}[ \mathrm{Ca}^{2+}] }  1 \bs 11
\label{eq:first_deactivation_a}
\end{equation}
\begin{equation}
0 \bs 00 \rates{k_3}{k_{-3}} 0 \bs 10  \xLeftrightarrow{ K_{1}[ \mathrm{Ca}^{2+}] }  1 \bs 10 \rates{k_2}{k_{-2}}  1 \bs 11
\label{eq:first_deactivation_b}
\end{equation}
\label{eq:first_deactivation}
\end{subequations}
The double arrow ($\Leftrightarrow$) indicates reversible binding and release of regulatory \Cm~by TnC. Activation is rightward; deactivation is leftward.

The kinetic schemes (Eqs.~\ref{eq:first_activation}, \ref{eq:first_deactivation}) simplify for experiments that involve rapid perturbations of [\Cm]~between the extremes of activation---full saturation and full desaturation. In the first path of activation (Eq.~\ref{eq:first_activation_a}), a large [\Cm] strongly favors $(1 \bs 00)$ over $(0 \bs 00)$, and \Cm~binding is rapid, so $(0 \bs 00)$ can be ignored on the millisecond and longer time scale.  In the second path of activation (Eq.~\ref{eq:first_activation_b}), rapid \Cm~binding to $(0 \bs 10)$ out-competes the deactivation of TnC through $k_{-3}$.  The first two steps can be approximated by a single transition with a forward rate constant $k_3$ and a backward rate constant $k'_{-3}$ \emph{that is small}. In the first path of deactivation (Eq.~\ref{eq:first_deactivation_a}), a low [\Cm] strongly favors $(0 \bs 11)$ over $(1 \bs 11)$, and the step of \Cm~release is assumed to not be rate limiting, so $(1 \bs 11)$ can be ignored. In the second path of deactivation (Eq.~\ref{eq:first_deactivation_b}), a low [\Cm] strongly favors $(0 \bs 10)$ over $(1 \bs 10)$. The step of \Cm~release is assumed to not be rate limiting, so $(1 \bs 10)$ can be ignored. Considering only the protein components ($\bs s_1 s_2$), the schemes for the two activation pathways simplify:
\begin{subequations}
\begin{equation}
 \bs 00 \rates{k_1}{k_{-1}}  \bs 10 \rates{k_2}{k_{-2}}  \bs 11    
\label{eq:activation_a}
\end{equation}
\begin{equation}
 \bs 00  \rates{k_3}{k'_{-3}} \bs 10  \rates{k_2}{k_{-2}} \bs 11    
\label{eq:activation_b}
\end{equation}
\label{eq:activation}
\end{subequations}
The two pathways of activation will relax with different rates from differences between $k_1$ and $k_3$ and between $k_{-1}$ and $k'_{-3} \simeq 0$. The relative flux between the two pathways of activation depends on the underlying rate constants for $K_0$ and $K_1$ as well as $k_1$, $k_{-1}$, $k_3$ and $k_{-3}$ in Eqs.~\ref{eq:first_activation}. Relative flux can not be predicted from the reduced models (Eqs.~\ref{eq:activation_a}).

In the two schemes for deactivation (Eq.~\ref{eq:first_deactivation}),  the protein system-states ($\bs s_1 s_2$) and the kinetic transition rates are identical.  The schemes are equivalent with respect to the measured FRET distance, which depends only on the states of the proteins. Considering only the protein components ($\bs s_1 s_2$), deactivation through either (or both) pathway(s) will precede according to the scheme
\begin{equation}
\bs 00 \rates{k_3}{k_{-3}} \bs 10 \rates{k_2}{k_{-2}}   \bs 11
\label{eq:deactivation}
\end{equation}
The assumed $\sim 10/\mu\mathrm{s}$ transition rate for \Cm~release from partially-activated TnC-TnI and fully-activated TnC-TnI can be relaxed to allow faster dissociation rates. This leads to more complicated kinetic schemes, where \Cm~can undergo one or more cycles of binding/release during activation or deactivation. In one such kinetic scheme of TnC-TnI activation, TnI activation $\sigma_2^+$ occurs after \Cm~has transiently dissociated from TnC:
\begin{equation}
0 \bs 00 \xLeftrightarrow{ K_{0}[ \mathrm{Ca}^{2+}] }  1 \bs 00 \rates{k_1}{k_{-1}} 1 \bs 10  \xLeftrightarrow{ K_{1}[ \mathrm{Ca}^{2+}] } 0 \bs10  \rates{k_2}{k_{-2}} 0 \bs 11  \xLeftrightarrow{ K_{2}[ \mathrm{Ca}^{2+}] } 1 \bs 11   
\label{eq:alt}
\end{equation}
This scheme extends the dominant pathway of activation (Eq.~\ref{eq:first_activation_a}) by including additional transitions of \Cm~binding and release, but it differs from Eq.~\ref{eq:first_activation_a} because $\sigma_2^+$ occurs when \Cm~is not bound. Because the transition rates $k_2$ and $k_{-2}$ are degenerate (i.e. independent of whether \Cm~is bound to TnC) the schemes  in Eq.~\ref{eq:first_activation_a} and Eq.~\ref{eq:alt} are kinetically equivalent with respect to the isomerization dynamics of TnC-TnI (Eq.~\ref{eq:activation_a}). Our current measurements can not distinguish between the activation schemes in Eq.~\ref{eq:first_activation_a} and Eq.~\ref{eq:alt} because their isomerization dynamics are given identically by Eq.~\ref{eq:activation_a}. The same arguments apply to other complicated kinetic schemes.

Eqs.~\ref{eq:activation} and \ref{eq:deactivation} are rationally reduced models for the isomerization dynamics of TnC and TnI during the activation and deactivation components of the signaling cycle when the TnC-TnI assembly is driven by step changes in \Cm~between fully saturating and fully de-saturating conditions. Eq.~\ref{eq:activation_b} represents a \emph{possible} accessory pathway of activation that may be experimentally resolved. The signaling cycle dynamics at the extremes of activation are determined by six rate constants: $k_{1}$, $k_{-1}$, $k_{2}$, $k_{-2}$, $k_{3}$, $k_{-3}$. Three structural configurations of TnC-TnI are accessible: $(\bs 00)$, $(\bs 10)$, and $(\bs 11)$.

\section{Results}

\subsection{Time-resolved FRET} 
\begin{figure}
\centering
\includegraphics[height=6cm]{\FIGS/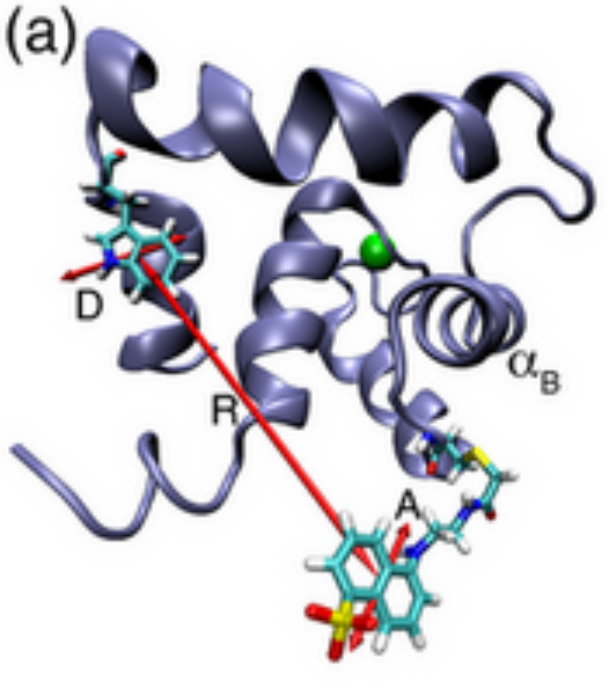}
\includegraphics[height=6cm]{\FIGS/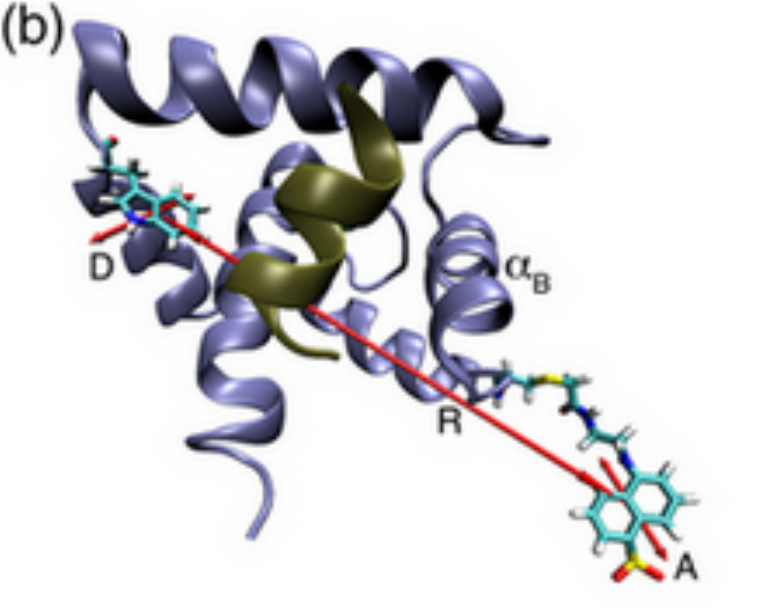}
\caption{
}
\label{fig:MD_sims}
\end{figure}

\begin{figure}
\centering
\includegraphics[scale=0.8]{\FIGS/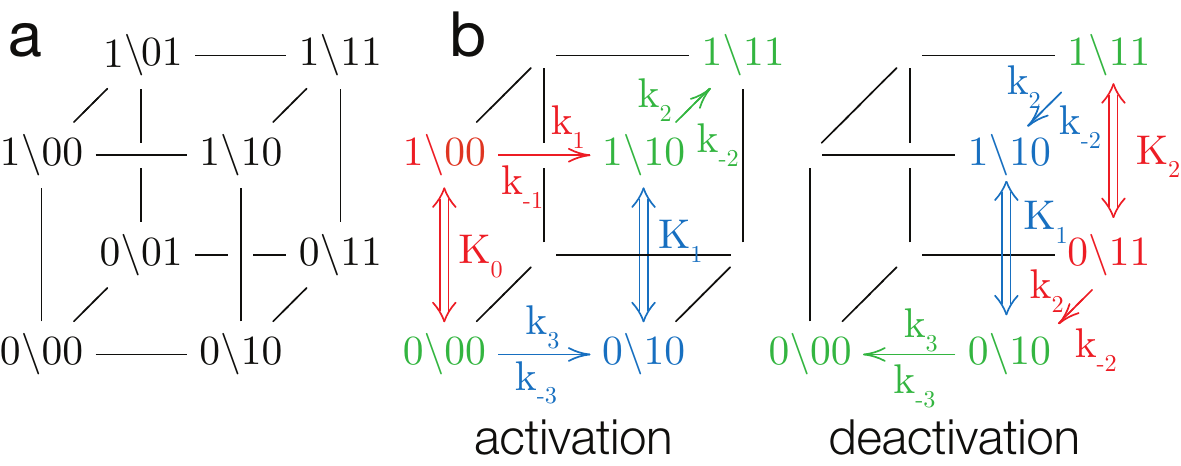}
\caption{\label{fig:reduced_system}
}
\label{fig:network_models}
\end{figure}

\begin{figure}
\centering
\includegraphics[scale=0.6]{\FIGS/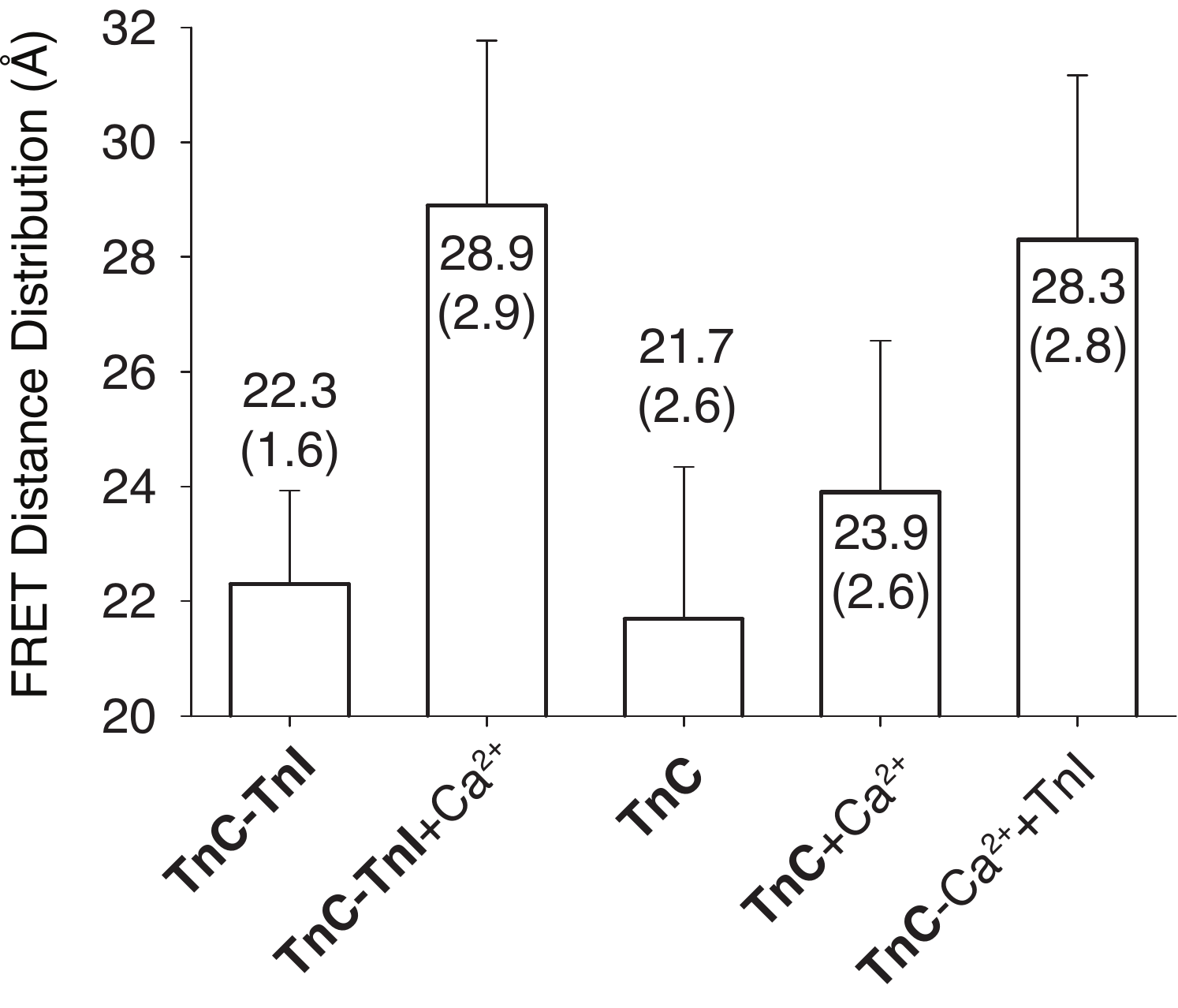}
\caption{
}
\label{fig:time_resolved}
\end{figure}

We employed a previously characterized FRET reporter system TnC(12W/51C*AEDANS)~\cite{Dong:1999mt} to follow structural changes during the signaling cycle. The FRET assay, shown in Fig.~\ref{fig:MD_sims}, is sensitive to TnC opening---the inter-helical rearrangement in TnC of helices B and C relative to the central helix D. The time-resolved decays of donor-only TnC(12W/51C) and donor-acceptor TnC(12W/51C*AEDANS) have been reported~\cite{Dong:1999mt} for isolated TnC and the TnC-TnI complex.  In this study we repeated and extended the measurements to include the addition of TnI to \Cm-saturated TnC. The multi-exponential decays (Supplemental Fig. 2) were fit to a static Gaussian distributed inter-probe distance model, as described~\cite{Dong:1999mt, Robinson:2004dg}. The recovered mean $\overline{r}$ and the standard deviation $\sigma$ of the inter-probe distance distribution are shown in Fig.~\ref{fig:time_resolved}. Here, but not in~\cite{Dong:1999mt}, background fluorescence was subtracted from the decays. Background subtraction resulted in small increases in $\overline{r}$ and a reduction in $\sigma$ of the distance distribution ($\sigma = 1.18 hw$) compared to previous measurements. $hw$ is the half width at half maximum that was used to quantify the breadth of the distribution in our previous work.

In Fig.~\ref{fig:time_resolved}, \Cm-induced structural changes are apparent in both isolated TnC and in the binary TnC-TnI complex. Starting with the preformed \Cm-depleted TnC-TnI complex, saturation with \Cm~caused an apparent mean 6.6 \AA~change in the inter-probe distance.  \Cm~saturation of the previously \Cm-depleted TnC-only sample caused a small 2.2 \AA~increase in the mean FRET distance. NMR measurements of the N-domain of cardiac TnC have shown a subtle 12\Deg~change in the C/D interhelical angle~\cite{Spyracopoulos:1997nt}. The observed 2.2 \AA~change in our FRET measurement, where the acceptor C51*AEDANS is in the short loop that connects the B and C helices---the B/C linker---is evidently sensitive this subtle structural change. Other possibilities for the apparent 2.2 \AA~inter-probe distance change, such as partial opening of TnC, or altered fluorophore mobility, can not be ruled out. Subsequent addition of TnI generated a larger 4.4 \AA~increase in the mean FRET distance. The sequential addition of \Cm~and TnI produced a distance change (6.6 \AA) of equal magnitude to the \Cm-induced change in the binary complex, although absolute distances were 0.6 \AA~less than for the preformed complex. These results demonstrate that activation is commutative---the order of \Cm~and TnI addition is not important. In the analysis that follows, we take 28.9 \AA~as the apparent mean inter-probe distance in the fully activated binary complex.   The standard deviation of the inter-probe distance distribution for the apo TnC-TnI complex is narrower than the other samples (by $\sim 1$ \AA).

\begin{figure}
\centering
$\begin{array}{l}
\multicolumn{1}{l}{\mbox{\bf (a)}} \\ [-0.53cm] 
\begin{minipage}[t]{0.5\linewidth}
	\centering
	\includegraphics[width=7cm]{\FIGS/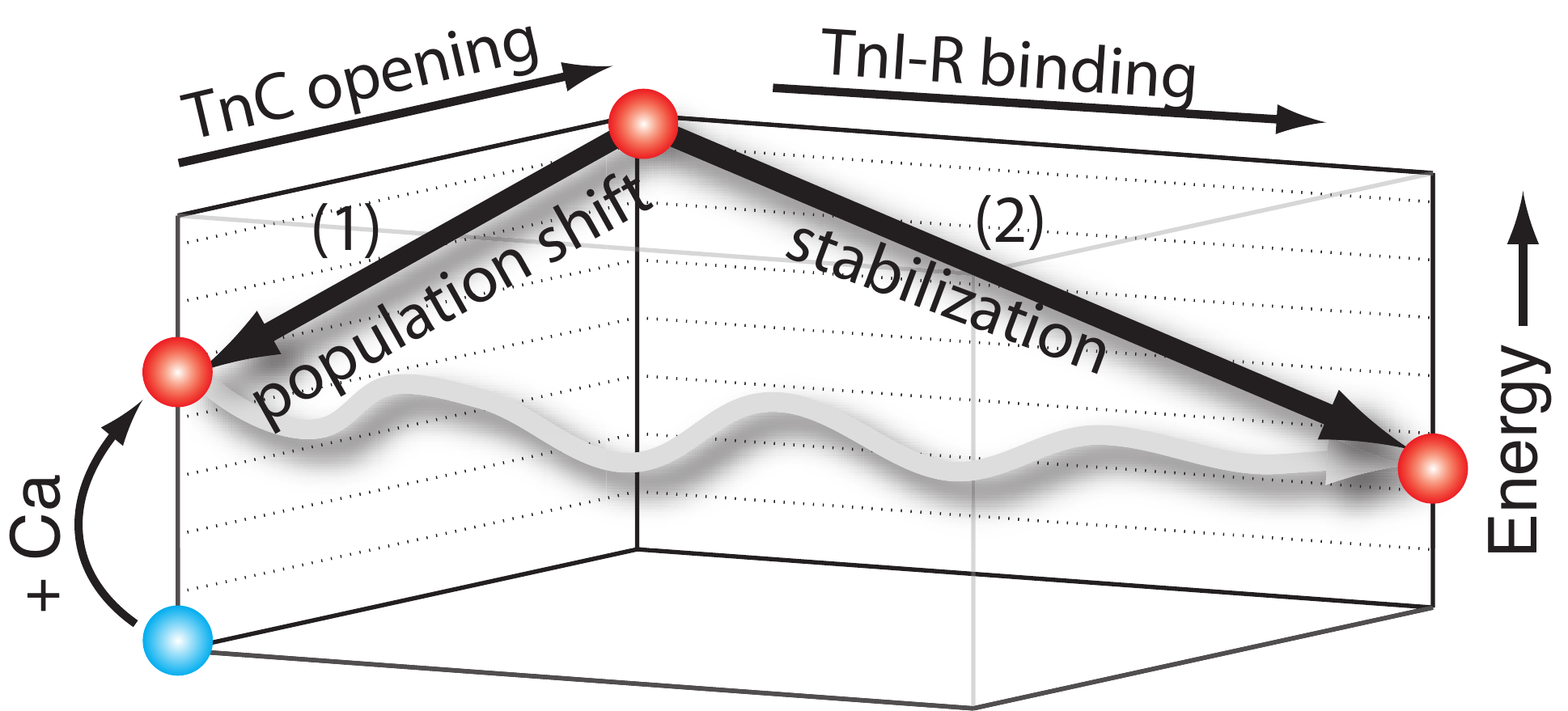}
	\includegraphics[width=4.5cm]{\FIGS/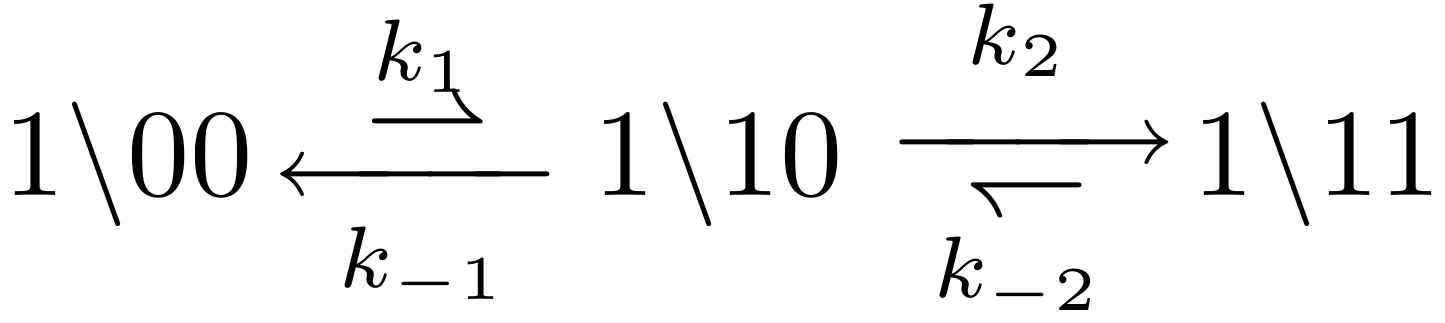}
\end{minipage} \\ [1.0cm] 
\multicolumn{1}{l}{\mbox{\bf (b)}} \\ [-0.53cm] 
 \begin{minipage}[t]{0.5\linewidth}
	\centering
	\includegraphics[width=7cm]{\FIGS/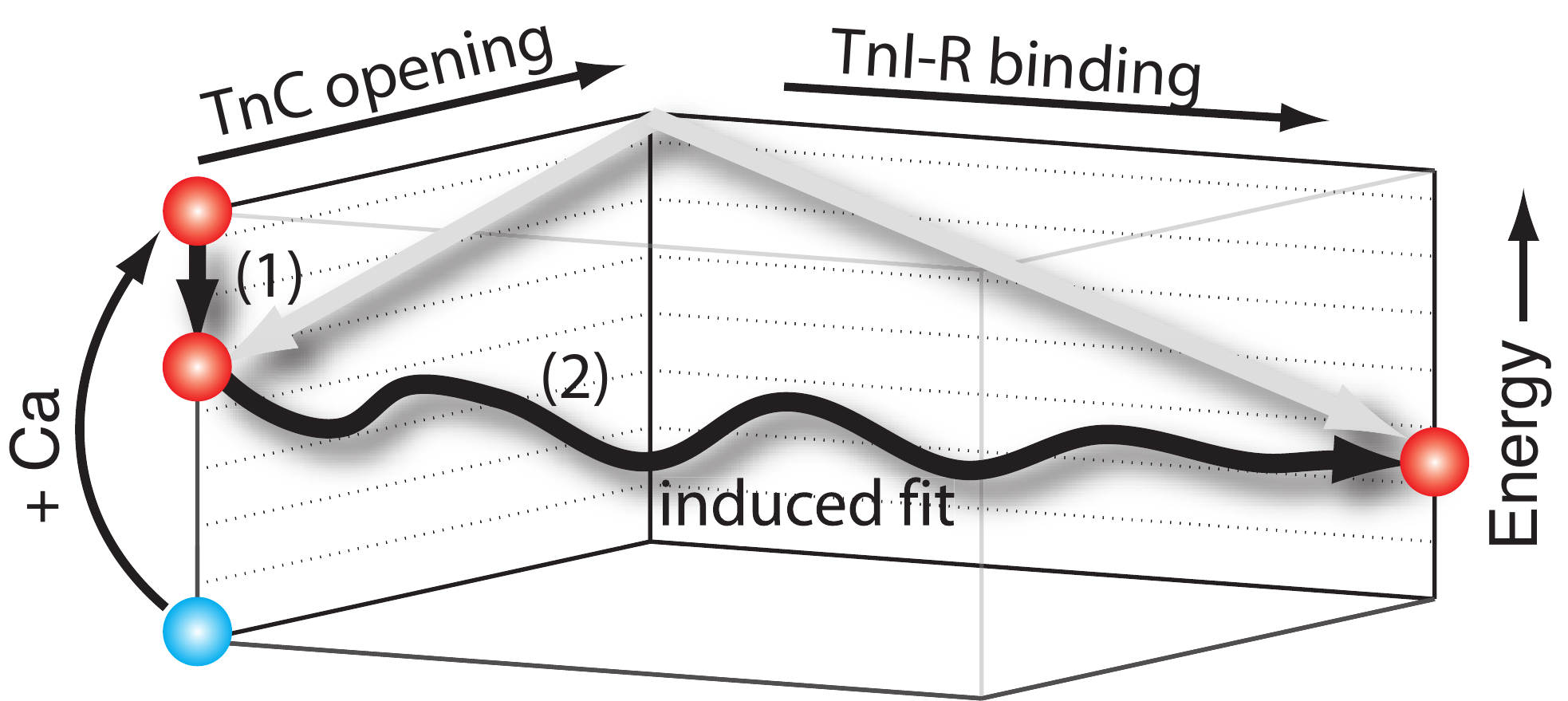} \\
	\includegraphics[width=4.5cm]{\FIGS/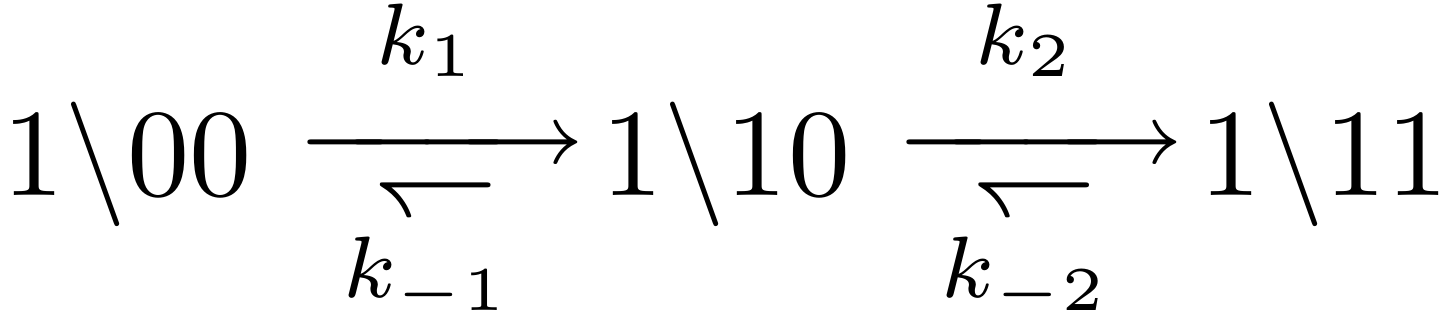}
\end{minipage}
\end{array}$
\caption{
}
\label{fig:models}
\end{figure}

Figure~\ref{fig:models} shows two mechanistic models of activation that can account for the small change in distance observed after \Cm~is added to isolated TnC and the large change in distance that is observed after the addition of  \Cm~to the binary TnC-TnI complex. The mechanistic models provide different structural interpretations of the allosteric transitions of TnC and TnI in the model of system dynamics (Fig.~\ref{fig:network_models}). In the population-shift-stabilization model (Fig.~\ref{fig:models}a), when TnI is not present, \Cm~binding to TnC fails to appreciably activate TnC because the transition is energetically unfavorable, $\mathcal{K}_1 \equiv k_{1} / k_{-1} < 1$.  Here, TnC activation is equated with TnC opening; that is, exposure of the hydrophobic pocket. Binding of TnI-R to the exposed hydrophobic pocket stabilizes an otherwise energetically unfavorable TnC opening event.  Without TnI, \Cm-bound TnC remains predominantly deactivated in the $(1 \bs 00)$ state.  Viewed kinetically, upon TnI addition most \Cm-bound TnC-TnI must first proceed through the rate limiting transition $k_{1}$ before becoming stabilized through the $k_2$ transition that out-competes the back reaction governed by $k_{-1}$: $k_2 > k_{-1}$. Combining these results, we find that both \Cm-induced and TnI-induced activation are rate-limited by $k_1$: $k_2 > k_{-1} > k_1$. If TnC-TnI switching is a population-shift-stabilization process, then the rate of \Cm-induced activation and the rate of TnI-induced activation will be \emph{equal}. The equality applies to the slowest relaxation rate of a multi-exponential decay. 

The induced-fit model interprets differently the small change in distance that is observed after adding \Cm~to isolated TnC compared to the large change in distance that is observed after adding  \Cm~to the binary TnC-TnI complex. In this model, \Cm~binding \emph{directly} and favorably activates TnC, $\mathcal{K}_1 > 1$, but the TnC activation event, $(\bs 00) \to (\bs 10)$, does not involve appreciable structural change (opening). Because of the large $\mathcal{K}_1$, \Cm-binding strongly shifts the distribution of states towards an energeticaly favorable ``\Cm-primed" state of TnC $(\bs 10)$ that we call the primed-closed state. TnI subsequently induces structural opening as TnI-R migrates between the B/C and D helices of TnC, which forces TnC to open. With TnC predominantly in the primed-closed state, activation flux in TnI-induced activation (TnI addition to \Cm-bound TnC) is rate-limited only by the opening step that is governed by $k_2$. The kinetic rate of a sequential reaction cannot exceed the kinetic rate of any component reaction. So, the rate of TnI-induced activation must be greater than or equal to the rate of \Cm-induced activation, which contains the isomerization steps of both TnC and TnI. For the observed \Cm-induced and TnI-induced distance changes (Fig.~\ref{fig:time_resolved}), the induced-fit mechanism allows the rate of TnI-induced activation to \emph{exceed} the rate of \Cm-induced activation. This is not the case for populaiton-shift-stabilization mechanism for which both TnI-induced and \Cm-induced activation are limited by $k_1$.

\subsection{Stopped flow kinetics}
To determine which of the two mechanistic models correctly describes activation of the TnC-TnI assembly, we performed a series of stopped-flow measurements of the activation and deactivation components of the signaling cycle. In each experiment, the mean time-dependent FRET distance was calculated from the transient donor probe fluorescence of independently measured concentration-matched preparations of donor-only and donor-acceptor samples, as described~\cite{Dong:2003pr}. Trp12 fluorescence in mock injected and actual samples were converted into time-dependent mean inter-probe distance using the mean distances that were recovered from the time-resolved FRET experiments (Fig.~\ref{fig:time_resolved}). 

\begin{figure}
\centering
$\begin{array}{l}
\multicolumn{1}{l}{\mbox{\bf (a)}} \\ 
\includegraphics[scale=0.6]{\FIGS/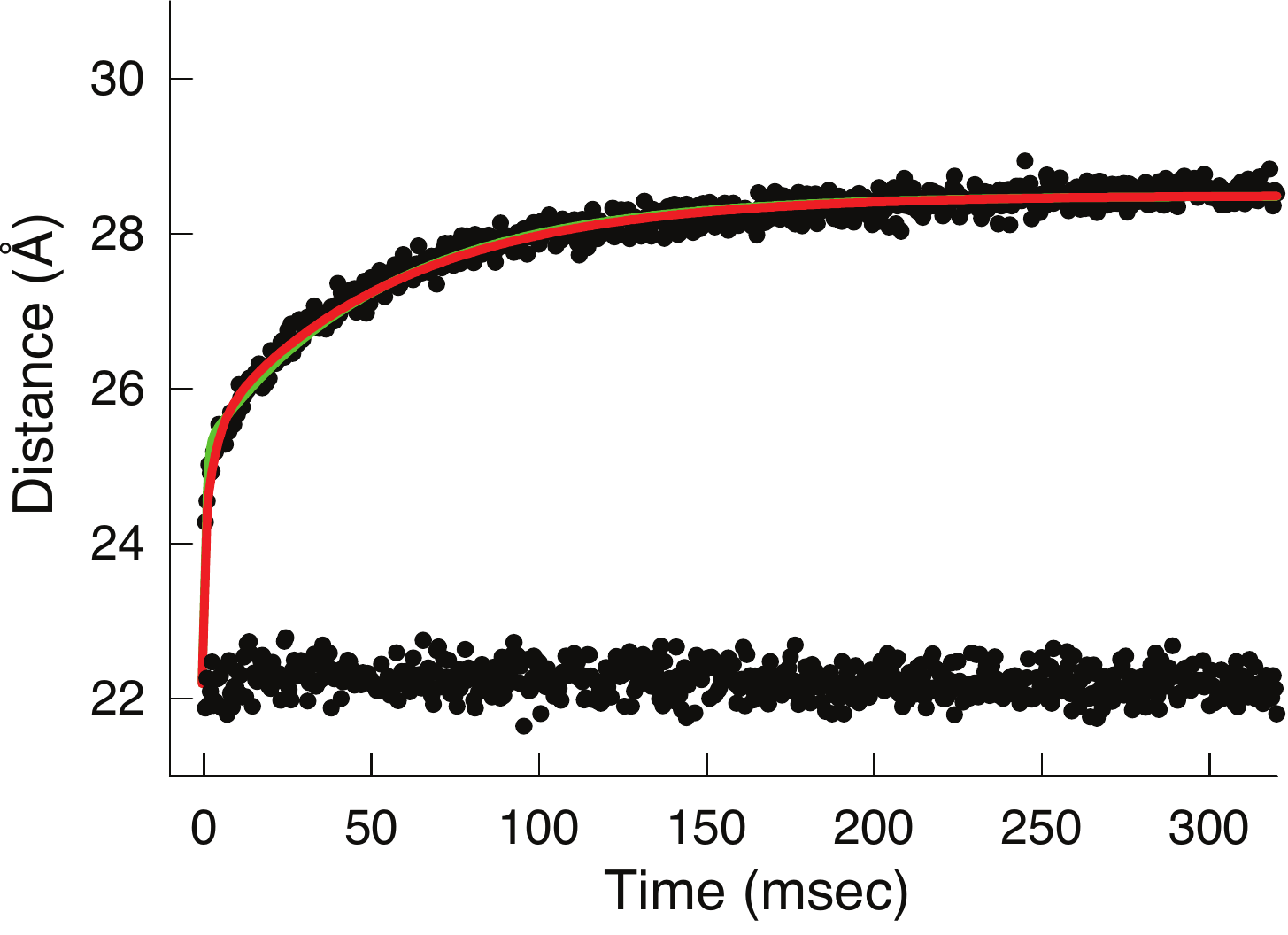} \\
\multicolumn{1}{l}{\mbox{\bf (b)}} \\ 
\includegraphics[scale=0.6]{\FIGS/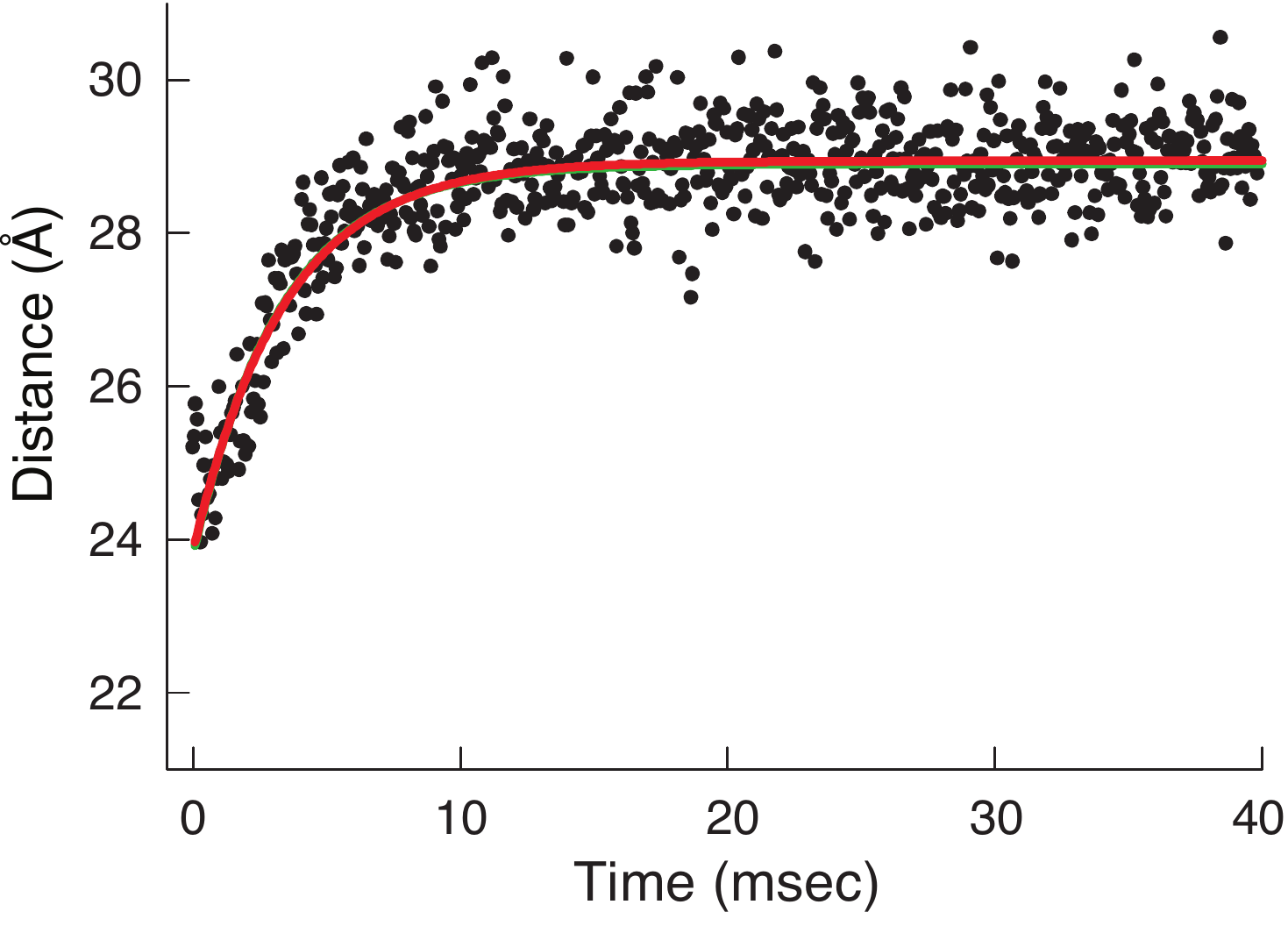} 
\end{array}$
\caption{
}
\label{fig:activation}
\end{figure}

Activation kinetics were monitored in two sets of stopped-flow measurements (Fig.~\ref{fig:activation}).  For \Cm-induced activation (Fig.~\ref{fig:activation}a), distance changes were obtained after rapidly mixing preformed binary TnC-TnI in a minimally \Cm-buffered solution (30 \uM~EGTA) with buffer containing sufficient \Cm~to saturate the sample (500 \uM). Empirically, the FRET relaxation is a bi-exponential process consisting of a very rapid transient ($\tau_1^{-1}$ = 1554 \sec, 48\%  amplitude) followed by much slower transition ($\tau_2^{-1}$ = 19 \sec, 52\%  amplitude). The reduced (unweighted) $\chi^2$, $\chi^2_R = \chi^2 / DOF$ was $2.9\times10^{-5}$. A fit to a three exponential function (not shown; $\chi^2_R$, $2.6\times10^{-5}$) did not provide a statistically significant improvement in the residuals, $F = \chi^2_{R(3 \exp)} /  \chi^2_{R(2 \exp)} = 0.87$. The 68\% confidence level is reached when $F<0.68$. 

For TnI-induced activation (Fig.~\ref{fig:activation}b) distance changes were obtained from samples of  \Cm-presaturated TnC that were rapidly mixed with a two-fold excess of isolated TnI. Rapid single exponential kinetics were observed ($\tau^{-1}$ = 305 \sec, $\chi^2_R$ = $7.0\times10^{-4}$). A two exponential fit (not shown; $\chi^2_R$ = $7.0\times10^{-4}$) did not improve the fitting, $F = 0.99$. We find that the rate of TnI-induced activation ($\tau^{-1}$ = 305 \sec) greatly exceeds the rate of \Cm-induced activation ($\tau_2^{-1}$ = 19 \sec), a rate that is sensitive to both \Cm- and TnI-activating steps. This finding is consistent with the induced-fit model, which, of the two mechanistic models, is the only one that allows the rate of cTnI-induced activation to exceed the rate of \Cm-induced activation. If the transition preceded through a population-shift-stabilization mechanism then the rate of TnI-induced activation would be rate-limited to 19 \sec. The rate of overall association between TnC and TnI is apparently much faster than the rate of TnI-induced activation. The observed rapid single exponential transient for TnI-induced activation indicates that saturation with \Cm~places TnC predominantly in the primed-closed state $(1 \bs 10)$ that precedes opening. From the \Cm-primed state, the TnC-TnI assembly rapidly undergoes an induced-fit opening through $k_{2}$. We conclude that the \Cm-priming step (with $k_{1} < 19$ \sec) is the rate limiting step in the activation of the TnC-TnI assembly. The \Cm-priming step is energetically favorable, $\mathcal{K}_{1} \gg 1$ because there is no slow phase in TnI-induced activation (from a population that must undergo a slow $k_1$ transition prior to opening through $k_2$). This implies that reverse reaction rate of \Cm-priming is very slow: $k'_{-1} = k_{1}/ \mathcal{K}_1 \simeq 0$ \sec. 

The allosteric model of activation (Fig.~\ref{fig:network_models}b; Eqs.~\ref{eq:activation_a},~\ref{eq:activation_b}) provides a straight-forward explanation of the two-exponential process of \Cm-activation.  The two exponential process is interpreted as the superposition of two populations of TnC-TnI, each relaxing through one of the two distinct pathways of activation. The values for $k'_{-1}$ and $k'_{-3}$ that were obtained above afford preliminary estimates of $k_1$ and $k_3$. For the first pathway of activation (Eq.~\ref{eq:activation_a}), the dominant eigenvalue of its rate matrix (Eq.~\ref{eq:activation_matrix}) is 
\begin{equation}
\tau^{-1} = (k_{1}k_{2} + k_{-1}k_{-2})/(k_{-1} + k_{2}).
\end{equation}
With $k_{-1} \simeq 0$, this simplifies to $\tau^{-1} \simeq k_{1} = 19$ \sec. For the second pathway of activation (Eq.~\ref{eq:activation_b}), the dominant eigenvalue of its rate matrix (Eq.~\ref{eq:deactivation_matrix}) is 
\begin{equation}
\tau^{-1} = (k_{3}k_{2} + k'_{-3}k_{-2})/(k'_{-3} + k_{2}).
\end{equation}
With $k'_{-3} \simeq 0$, this simplifies to $\tau^{-1} \simeq k_{3} = 1554$ \sec. The activation models provide a simple expression for the observed rate for TnI-induced activation,
\begin{equation}
\tau^{-1} = k_{2} + k_{-2}.
\label{eq:TnI_eigen}
\end{equation}
Additional information needed to resolve $k_2$ and $k_{-2}$ is provided by the kinetics of deactivation.

Deactivation of the TnC-TnI assembly was monitored following rapid mixing of a solution containing the \Cm~chelator EGTA (2 mM) with a \Cm-saturated sample of TnC-TnI (Fig.~\ref{fig:deactivation}). The relaxation transient was empirically fit as a single exponential decay process ($\tau^{-1}$ = 125 \sec, $\chi^2$ = $1.7\times10^{-4}$). A two exponential fit (not shown) did not statistically improve the fitting ($F = 0.76$). Because the scheme for deactivation is kinetically linked with the scheme for TnI-induced activation (Eqs.~\ref{eq:activation}, \ref{eq:deactivation}), the information in the two experiments can be combined to resolve $k_{2}$ and $k_{-2}$. The dominant eigenvalue of the scheme of deactivation (Eq.~\ref{eq:deactivation}) is
\begin{equation}
\tau^{-1} = (k_{3}k_{2} + k_{-3}k_{-2})/(k_{-3} + k_{2}).
\end{equation} 
For the system to deactivate upon the release of \Cm, the system must preferentially exit the primed-closed state $(\bs 10)$ to the left; this implies that $k_{-3} \gg k_{2}$.  The expression for the observed rate simplifies to
\begin{equation}
\tau^{-1} = \mathcal{K}_3 k_{2} + k_{-2},
\label{eq:deact_eigen}
\end{equation}
where $\mathcal{K}_{3} \equiv k_{3}/k_{-3}$. Eqs.~\ref{eq:TnI_eigen} and~\ref{eq:deact_eigen} explain why the rate of relaxation for deactivation (125 \sec) is slower than the rate of TnI-induced activation (305 \sec). In Eq.~\ref{eq:deact_eigen}, $k_2$ is pre-multiplied by a $\mathcal{K}_3 < 1$. $\mathcal{K}_3$ must be less than unity for TnC to swich off during deactivation. Assuming that $\mathcal{K}_{3} = 0.1$ we obtain estimates for $k_2$ (200 \sec)~and $k_{-2}$ (105 \sec). The backward reaction rate $k_{-2}$ approaches the forward reaction rate $k_{2}$, making induced-fit TnC opening highly reversible.  Under saturating \Cm~(15 \Celcius), the system is an approximately 2:1 mixture of rapidly inter-converting fully activated (open) and partially activated (primed-closed) species.  The finding that the TnC-TnI does not fully activate upon \Cm~binding implies that the mean FRET distance of the \Cm~saturated TnC-TnI assembly (Fig.~\ref{fig:time_resolved}) is a weighted mixture of the FRET distances from primed-closed and open populations.

\subsection{Global analysis of equilibrium and transient FRET distances}
\begin{figure}
\centering
\includegraphics[scale=0.6]{\FIGS/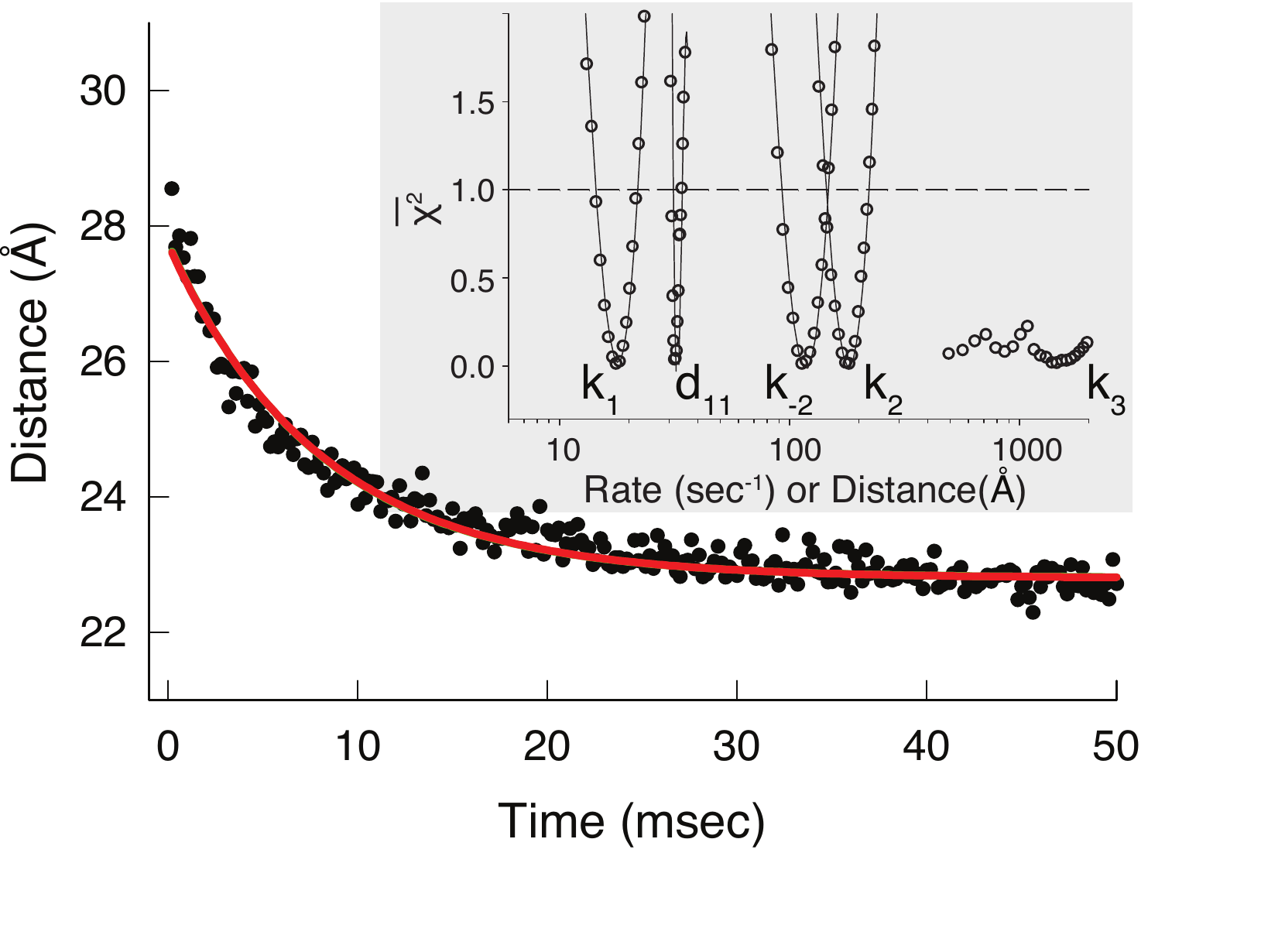}
\caption{
}
\label{fig:deactivation}
\end{figure}

Initial estimates for the rate parameters that govern the model of TnC-TnI allostery (Eqs.~\ref{eq:activation} and~\ref{eq:deactivation}) were obtained from the observed relaxation rates in the stopped-flow FRET measurements and assumed equilibrium constants. The structural information provided by FRET was ignored. To realize the full potential of the experimental information, we performed a global analysis of the three stopped-flow FRET measurements. The main task of the global fitting was to resolve the species-associated FRET distances and the microscopic transition rates that cause system-state populations to evolve following perturbation of the system. The results of the global fitting are shown in Figs.~\ref{fig:activation} and~\ref{fig:deactivation}. Recovered values of the model parameters $k_{1}$, $k_{2}$, $k_{-2}$, $k_{3}$, $d_{\bs 00}$, and $d_{\bs 11}$ are given in  Table~\ref{tbl:params}. To provide stable convergence, $k_{-1}$, $k_{-3}$ and $d_{\bs 10}$ were assigned fixed values. The FRET distance $d_{\bs 10}$ was fixed at 23.9 \AA, the mean FRET distance of the \Cm-saturated TnC-only sample from the time-resolved measurements. Parameters $k_{-1}$ and $k_{-3}$ were assigned fixed values of 1.0 \sec~and 31000 \sec~because they were, respectively, too slow or too fast to be resolved by the data. Values were assigned based on the preliminary estimates of $k_{1}$ and $k_{3}$, the requirement that the primed-closed state is favored over the closed state when \Cm~is bound, $\mathcal{K}_1> 1$, and the requirement that closed state is favored over the primed-closed state when \Cm~is not bound, $\mathcal{K}_3 < 1$.

The ability of the experimental information to specify values for the model parameters is quantitated by the recovered precision of the model parameters. The precision of the parameters $k_{1}$, $k_{2}$, $k_{-2}$, $k_{3}$, and $d_{\bs 11}$ were determined by projecting the $\chi^2$ hyper-surface (measure of deviation between model and data as a function of parameter values) along the individual parameter axes in a series of 1-D adiabatic grid searches (Fig.~\ref{fig:deactivation}, inset; Table~\ref{tbl:params}).  The grid search results are reported as a normalized $\chi^{2}$
\begin{equation}
\overline{\chi^2}(x) = \frac{ \chi^{2}(x) - \chi^{2}_{min} } { \chi^{2}_{tgt} - \chi^{2}_{min} },
\label{eq:chi_sqr}
\end{equation}
where $\chi^{2}_{min}$ is the $\chi^{2}$ at the converged minimum. $\chi^{2}(x)$ are $\chi^{2}$ values at grid search points. The target $\chi^{2}$, $\chi^{2}_{tgt} = \chi^{2}_{min} (1 + \alpha F(\delta) )$, is a function of the  F-statistic   $F(\delta)$, where $\delta$ is the confidence level (taken as 68\%), and $\alpha \simeq 1$ accounts for the loss of one degree of freedom in the grid search. Intersection of $\overline{\chi^2}$ with unity (dashed line) provides the upper and lower 68\% (one standard deviation) error estimates of the parameter. In general, parameter error is non-Gaussian, and it is preferable to express parameter precision using upper and lower limits that non-symetrically bracket the 68\% confidence interval.  The values of parameters $k_{1}$, $k_{2}$, $k_{-2}$, and $d_{\bs 11}$ are determined with high precision (Fig.~\ref{fig:deactivation}, inset).  The parameter $k_{3}$ is poorly resolved since the $\overline{\chi}$ curve for $k_{3}$ is shallow; it also has more than one local minimum. This is expected since $k_3 = 1618$ \sec~exceeds the resolution of the stopped flow instrument (1/1.8 ms dead time = 555 \sec). The value of the species associated distance $d_{\bs 00}$ depends on the assumed value for $k_{-1}$.  Since the accuracy of $d_{\bs 00}$ is conditional, its precision was not determined. 

The refined values of parameters $k_{2}$ (181 \sec)~and $k_{-2}$ (117 \sec)~confirmed the preliminary observation that \Cm-bound TnC-TnI is in dynamic equilibrium between the primed-closed and open conformations.  The micro-equilibrium constant  of TnI-facilitated TnC opening is rigorously determined, $\mathcal{K}_2 \equiv k_{2}/k_{-2} = 1.55$ (15 \Celcius).  Estimates for the micro-association constants that govern \Cm-induced TnC priming (isomerization of $s_1$), $\mathcal{K}_1 \equiv k_1/k_{-1} = 18.0$ and $\mathcal{K}_3 \equiv k_3 / k_{-3} = 0.052$, are less reliable because $k_{-1}$ and $k_{-3}$ were assigned fixed values; they are, nevertheless, consistent with the data. The mean observed FRET distance of \Cm-saturated TnC-TnI from the time-resolved measurements (28.9 \AA ) is a 1.55:1 mixture of species-associated inter-probe distances for the primed-closed state $d_{\bs 10}$ (mean, 23.9  \AA)~and the open state $d_{\bs 11}$ (mean, 32.2 \AA).  The mean inter-probe distance change for TnC opening is 8.3 \AA.

\subsection{FRET-\Cm~titration}
From free energy conservation, the three micro-equilibrium constants that govern the protein isomerizations $\mathcal{K}_1$, $\mathcal{K}_2$, $\mathcal{K}_3$, along with any of the three equilibrium constants for \Cm~binding to closed ($\bs 00$), primed-closed ($\bs 10$), or open ($\bs 11$) system states (respectively, $K_0$, $K_1$ and $K_2$) provide a  complete parametrization of the free energies $\set{G_{S_i}}$ of the relevant system-states in the macroscopic free energy landscape $\set{S_i, G_{S_i}}$ of the TnC-TnI assembly. To complete the thermodynamic parametrization of the TnC-TnI landscape, we performed an equilibrium \Cm~titration of the FRET distance~(Fig.~\ref{fig:titration}a). For reference, the data were empirically fit to the Hill equation, $R = R_{\mathrm{min}} + (R_{\mathrm{max}} - R_{\mathrm{min}}) (1 + 10^{n(pCa - pCa_{50})})^{-1}$. The recovered parameters ($n=1.18$, $pCa_{50} =5.82$) are consistent with previous measurements~\cite{Robinson:2004dg, Dong:2003pr}. Through the Boltzmann equation (Eq.~\ref{eq:Boltzmann}), the $G_{S_i}$ determine how the system will equilibrate among all system-states $S_i$ at a given [\Cm].  As discussed in Appendix B, the free energies of the ligand-unbound species depend linearly on the \Cm~chemical potential, $\mu = - \ln (10) R_{B}T pCa$: $G_{0 \bs s_1 s_2} = G_{0 \bs s_1 s_2}^{\circ} + \mu$. The free energies of the ligand-bound species are independent of $\mu$: $G_{1 \bs s_1 s_2} = G_{1 \bs s_1 s_2}^{\circ}$. The relative free energies of the ligand-bound species are calculated from $k_{2}$, $k_{-2}$, $k_{3}$, and $k_{-3}$ using 
\begin{equation}
\beta \Delta G_{i}^{\circ} = -\ln(k_{i}/k_{-i}),
\label{eq:micro_DB}
\end{equation}
where $\beta = 1/ R_{B}T$ and we assign $\beta G^{\circ}_{1 \bs 00}$ = 0. From the rate constants in Table~\ref{tbl:params}, we obtain $\beta G_{1 \bs s_1 s_2} = (0, -2.89, -3.33)$. The relative free energies of the ligand-unbound species ($\beta G_{0 \bs s_1 s_2}$) were calculated from $k_{1}$, $k_{-1}$, $k_{2}$, and $k_{-2}$ after assigning $G^{\circ}_{0 \bs 00} = \mu'$, where  $\mu'$ is a chemical potential offset that specifies when $G^{\circ}_{0 \bs 00} =  G^{\circ}_{1 \bs 00}$. From Eq.~\ref{eq:micro_DB} and the rates in Table~\ref{tbl:params}, we obtain, $\beta G_{0 \bs s_1 s_2} = (0, 2.95, 2.51) + \beta \mu + \beta  \mu'$. A fit to the \Cm-FRET titration data using Eqs~\ref{eq:marginal},~\ref{eq:R},~\ref{eq:Boltzmann} (Fig.~\ref{fig:titration}a) provided the value for $\beta \mu'$ (9.56)---the remaining parameter needed to parametrize the landscape (Fig.~\ref{fig:titration}b). The datum is sufficient to determine the remaining micro-equilibrium constants $K_0$, $K_1$, $K_2$ that respectively determine the affinity of \Cm~for the system states $(\bs 00)$, $(\bs 10)$, $(\bs 11)$ (c.f. Fig.~\ref{fig:network_models}). $K_0 = \exp(\beta \mu')= \exp(9.56) =1.4 \times 10^4$ M$^{-1}$. From conservation of free energy and considering the energy changes corresponding to $\mathcal{K}_1$ and  $\mathcal{K}_3$, we obtain $K_1 = \exp(9.56+2.95+2.89)=4.9 \times 10^6$ M$^{-1}$. We conclude that the primed-closed state has  $K_1/K_0 \cong 300$-times higher affinity for \Cm~than the closed state. Due to nearest-neighbor-limited influence, there is no relative free energy difference between the \Cm-bound and \Cm-unbound surfaces for the opening transition, so $K_2 = K_1 = 4.9 \times 10^6$ M$^{-1}$. $K_{\mathrm{app}}$, the net (or apparent) affinity of \Cm~for the regulatory site on TnC in the TnC-TnI assembly can be calculated from the recovered landscape $\set{S_i, G_{S_i}}$ using Eqs.~\ref{eq:O},~\ref{eq:Ca-bind} and~\ref{eq:Boltzmann} (see Discussion section). $K_{\mathrm{app}} = 10^{pK_{50}} = 10^{5.78} = 6.0 \times 10^5$, is a function of the entire landscape $\set{S_i, G_{S_i}}$. The $pCa_{50}$ (5.82) from the fit to the Hill equation closely corresponds to the recovered $pK_{50}$ (5.78).

\subsection{Temperature-dependent kinetics}
To further parametrize the kinetics and thermodynamics of the opening step in activation, we repeated the stopped flow measurements (Figs.~\ref{fig:activation} and~\ref{fig:deactivation}) for a range of temperatures $T = \set{3.7, 8, 12, 15, 16, 20}$ \Celcius. TnI-induced activation was single exponential at all temperatures with recovered $\tau^{-1} = \set{222,254,299,305, 338,362}$~\sec. The kinetic transients were globally analyzed to obtain $k_{2}(T)$ and $k_{-2}(T)$~(Fig.~\ref{fig:Arrhenius_plot}). Noise increased with temperature (data not shown), and the global analysis failed to converge for 20 \Celcius.  The backward rate constant $k_{-2}$ showed upward deviation from a logarithmic dependence with temperature: $k_{-2} = \set{49, 61, 83, 103, 117}$ \sec~for $T = \set{3.7, 8, 12, 15, 16}$ \Celcius. The forward rate constant $k_2$ showed downward deviation from a logarithmic dependence on temperature: $k_2 = \set{175, 196, 218, 191, 181}$ \sec~for $T = \set{3.7, 8, 12, 15, 16}$ \Celcius. These deviations from Arrhenius Law behavior can not be attributed solely to a temperature-dependent energy (or enthalpy) of activation, which would cause $k_{2}$ and $k_{-2}$ to deviate in the same direction.

In an Arrhenius analysis of a \emph{reversible} first order process, the logarithm of the observed relaxation rate $\tau^{-1} = k_{2} + k_{-2}$ is plotted against inverse temperature $\beta =1/R_{B}T$. The slope (see Appendix A),
\begin{equation}
\pd{\ln(\tau^{-1})}{\beta} =  - \delta^{\ddagger} + \frac{k_{-2}}{k_{2} + k_{-2}}\Delta H,
\label{eq:slope_main}
\end{equation}
depends on the enthalpy of activation of the forward reaction $\delta^{\ddagger}$ and the net enthalpy change of opening, $\Delta H$. In the derivation of Eq.~\ref{eq:slope_main}, $\Delta H$ and the net entropy of opening $\Delta S$ were assumed to be temperature independent. The second term in Eq.~\ref{eq:slope_main}, which arises from reaction reversibility, can produce curvature in an Arrhenius plot. To recover the enthalpy $\Delta H$ and entropy $\Delta S$ of the opening process, we performed a van't Hoff analysis (Fig.~\ref{fig:Vant_Hoff_plot}) of the recovered temperature-dependent $k_2(T)$ and $k_{-2}(T)$ by fitting to 
\begin{equation}
f(\set{\Delta H,\Delta S};\beta) =  \frac{k_{2}(\beta)}{k_{-2}(\beta)} = e^{\Delta S/ R} e^{-\beta \Delta H},
\label{eq:Vant_Hoff_fit}
\end{equation}
where $f(\set{\cdot};\beta)$ indicates that parameters $\set{\cdot}$ are being fit to a function $f$ that depends on $\beta$. The rates at 16 \Celcius~were excluded from the fitting. Opening is exothermic ($\Delta H$, -33.4 kJ/mol) and is balanced by a loss of entropy ($\Delta S$, -0.110 kJ/mol/K). Curvature in the data points in Fig~\ref{fig:Vant_Hoff_plot} suggested that $\Delta H$ and $\Delta S$ undergo temperature-dependent change over the experimental temperature range. The data were refit to Eq.~\ref{eq:Vant_Hoff_fit} using temperature-dependent expressions for enthalpy and entropy, 
\[
\left.
\begin{array}{ccc}
\Delta H & = & \Delta H^o + \Delta C_p \left( T - T_0\right) \\
\Delta S & = & \Delta S^o + \Delta C_p \ln \left( T/ T_0 \right)
\end{array},
\right.
\]
where $\Delta H^o = \Delta H(T_0)$, $\Delta S^o = \Delta S(T_0)$, and the heat capacity difference between the open and closed-primed states $\Delta C_p$ is assumed to be temperature-independent. Fitting (Fig.~\ref{fig:Vant_Hoff_plot}, dashed line) produced maximum likelihood estimates for $\Delta H^o$ (-33.1 KJ/mol), $\Delta S^o$ (-0.108 KJ/mol/K), $\Delta Cp$ (-7.6 kJ/mol/K), and $T_0$ (282 K). The fit suggests that the temperature dependence of $K_2 = k_2 / k_{-2}$ can be attributed to a large negative $\Delta C_p$. The recovered $\Delta H^o$ and $\Delta S^o$ are within 2\% of the recovered temperature-independent $\Delta H$ and $\Delta S$.

To recover $\delta^{\ddagger}$ and the entropy-adjusted barrier crossing attempt frequency  $\nu_{\mathrm{adj}} \equiv \nu \exp \left( \sigma^{\ddagger} / R\right) $ ($\nu$ is the attempt frequency, and $\sigma^{\ddagger}$ is the entropy of activation of the forward reaction) the observed temperature-dependent rates $\tau^{-1} = k_{2} + k_{-2}$ for TnI-induced activation (c.f. Fig.~\ref{fig:activation}b), were fit to (see Appendix A)
\begin{equation}
f\left( \set{\nu_{\mathrm{adj}},\delta^{\ddagger}};\beta \right) = \tau^{-1}(\beta) =\nu_{\mathrm{adj}} e^{-\beta \delta^{\ddagger}} \left[ 1 + e^{\beta \Delta H -\Delta S / R} \right]
\label{eq:Arrhenius_fit}
\end{equation}
using the temperature-independent $\Delta H$ and $\Delta S$ obtained from the van't Hoff analysis. The fitting (Fig.~\ref{fig:Arrhenius_plot}, solid line) provided values for $\nu_{\mathrm{adj}}$ ($1.8 \times 10^4$ \sec) and $\delta^{\ddagger}$ (10.6 kJ/mol). Curvature in the slope is caused by reaction reversibility, a non-zero $\Delta H$, and shifting balance of forward $k_{2}$ and reverse $k_{-2}$ reaction rates. 
The data are well fit by Eq.~\ref{eq:Arrhenius_fit}, which does not include temperature-dependent changes in $\Delta H$ and $\Delta S$. Using a more complicated expression with temperature-dependent $\Delta H$ and $\Delta S$ is unlikely to improve the fitting results.

The combined results of the van't Hoff and Arrhenius analyses provide the reaction enthalpy $\Delta H$ (-33.4 kJ/mol), the reaction entropy $\Delta S$ (-0.110 kJ/mol/K), the enthalpy of activation of the forward reaction $\delta^{\ddagger}$ (10.6 kJ/mol), and the entropy-adjusted barrier crossing attempt frequency for the forward reaction $\nu_{\mathrm{adj}}$ ($1.8 \times 10^4$ \sec). From these values we obtain the enthalpy of activation of the reverse reaction, $\delta_{\mathrm{R}}^{\ddagger} \equiv  \delta^{\ddagger}-\Delta H$ (44.0 kJ/mol) and the entropy-adjusted barrier crossing attempt frequency for the reverse reaction, $\nu_{\mathrm{adj, R}} \equiv \nu \exp \left( \sigma_{\mathrm{R}}^{\ddagger} / R\right)$ ($1.0 \times 10^{10}$ \sec). $\nu_{\mathrm{adj, R}}$  is a function of the entropy of activation for the reverse reaction, $\sigma_{\mathrm{R}}^{\ddagger} \equiv \sigma^{\ddagger} - \Delta S$, calculated using $\nu_{\mathrm{adj, R}} = \nu_{\mathrm{adj}} \exp \left( - \Delta S / R\right) $. To confirm the fitting and calculations, $k_2 = \nu_{\mathrm{adj}}\exp(-\beta \gamma^{\ddagger})$ and $k_{-2} = \nu_{\mathrm{adj, R}}\exp(-\beta \delta_{\mathrm{R}}^{\ddagger})$ were calculated from the recovered parameters $\nu_{\mathrm{adj}}$, $\nu_{\mathrm{adj, R}}^{\ddagger}$, $\gamma^{\ddagger}$, $\gamma_{\mathrm{R}}^{\ddagger}$. In Fig.~\ref{fig:Arrhenius_plot}, the calculated rates (dashed lines) are plotted  along with the observed $k_2(T)$, $k_{-2}(T)$ (squares). The recovered thermo-kinetic parameters for the opening/closing transition are summarized in Table~\ref{tbl:landscape_params}.

\section{Discussion}
A previously characterized FRET reporter system TnC(12W/51C*AEDANS)~\cite{Dong:1999mt} was used to probe the structural kinetics of the cardiac TnC-TnI assembly during rapid activation and deactivation. In stopped-flow kinetic measurements, FRET provides a meaningful measure of inter-probe distance in addition to relaxation rates. The data were analyzed in terms of a new non-equilibrium mesoscopic (coarse-grained) model of allosteric transitions in the TnC-TnI assembly. The model captures the functional dynamics of the TnC-TnI assembly in two kinetically linked allosteric models for the activation and deactivation stages of the signaling cycle.

\subsection{Dynamic conformational equilibrium}
Our measurements provide structural and kinetic evidence for the existence of a dynamic equilibrium of macrostates in the TnC-TnI assembly when \Cm~is bound and when \Cm~is not bound to the regulatory site (loop II) of TnC.  These results provide mechanistic insight into the phenomenon of ``incomplete myofilament activation by \Cm"~\cite{Lehrer:1994yj, Robinson:2004dg}. In addition, the results provide, apparently, the first reported evidence of incomplete deactivation of cardiac troponin. The structural transition from the primed-closed state to the open state is governed by the forward rate constant $k_{2}$ ($181 \pm \set{145, 222}$ \sec) and backward rate constant $k_{-2}$ ($117 \pm \set{93, 148}$ \sec). Parameter precision $\pm \set{le, ue}$ is quantified by the lower and upper estimates for the 68\% confidence interval. We conclude that \Cm-bound TnC-TnI rapidly inter-converts between the primed-closed and open structural conformations, spending $\sim$65\% of the time in the open state ($T$ = 15 \Celcius). This finding is consistent with previously reported energetic measurements of \Cm-dependent changes in the affinity of TnC for TnI~\cite{Liao:1994un} and by high resolution structural studies of \Cm-saturated TnC~\cite{Spyracopoulos:1997nt, Takeda:2003lr}. The affinity constants  ($K_a^{\mathrm{Mg}}$, $K_a^{\mathrm{Ca}}$) of native TnI for TnC under \Mm-saturated and \Cm-saturated conditions (400mM KCl, 20 C) have been reported by Liao et. al.~\cite{Liao:1994un}. The ratio $\mathcal{K}_I = K_a^{\mathrm{Ca}}/ K_a^{\mathrm{Mg}}$ provides an estimate of the \Cm-sensitive component of TnC-TnI interaction. Equating the \Cm-sensitive TnC-TnI interaction with the TnC opening transition, we calculate, $\mathcal{K}_I = 127.0 \times 10^6\,\mathrm{M}^{-1}/ 41.7 \times 10^6\,\mathrm{M}^{-1} = 3.04$ (400 mM KCl, 20 C). This is in general agreement with the isomerization constant for TnC opening obtained in this study, $\mathcal{K}_2 = k_2/k_{-2} = 1.55$ (200 mM KCl, 15 C). 

Our measurements and analysis indicate that the TnC opening causes an 8.3~\AA~change in the mean inter-probe distance.  This distance change is statistically indistinguishable from the inter-probe distance change (9.2~\AA) calculated from the NMR-derived \Cm-bound primed-closed structure (PDB ID code 1AP4)~\cite{Spyracopoulos:1997nt} and X-ray crystallography-derived \Cm-bound open structure (PDB ID code 1J1E)~\cite{Takeda:2003lr} structure (JMR, manuscript in preperation). The inter-probe distance was obtained by modifying the high-resolution primed-closed and open structures in-silico to incorporate the donor and acceptor FRET probes (Fig.~\ref{fig:MD_sims}) and performing all-atom molecular dynamics simulations to sample the conformations of the FRET probes.

When regulatory \Cm~is not bound to TnC, the TnC-TnI assembly may exist in the partially activated primed-closed state in addition to the closed state. Activation through the accessory pathway in our allosteric model (Fig.~\ref{fig:models}b (blue); Eq~\ref{eq:first_activation_b}) occurs through \Cm-binding to a pre-existing population of TnC-TnI in the primed-closed state. Our kinetic measurements detect activation through this pathway. Solzin et al., using an environmentally sensitive fluorophore to monitor  activation kinetics, also detected a rapid transient during \Cm-activation~\cite{Solzin:2007qy}. They interpreted this transient as an additional step that precedes the rate-limiting (\Cm-priming) step in a serial reaction scheme for \Cm-activation. From the FRET-provided distance information in our measurements, we found that the rapid \Cm-induced transient (1554 \sec) involves substantial (48\% of 6.6 \AA) distance change. This large distance change is associated with opening (not priming). We have shown that opening temporally \emph{follows} the slow priming transition. The accessory pathway of activation thus bypasses the slow priming step in activation through \Cm~binding to a pre-existing population of TnC-TnI in the primed-closed state. The accessory pathway of activation suggests  that the TnC-TnI assembly does not completely deactivate to the closed conformation when regulatory \Cm~is not bound.

\subsection{Mesoscopic modeling}
Our mesoscopic model of TnC-TnI allostery (Fig.~\ref{fig:network_models}b) recognizes transition rate degeneracy due to nearest-neighbor-limited influence, and it recognizes the kinetic linkages between activation and deactivation stages of the signaling cycle. Transition rate degeneracy arises from the complex nature of the system. Simple systems such as chemical networks---networks of chemical reactions of diffusing molecules---do not exhibit transition rate degeneracy. The model shows, for example, that the elementary transition rate parameters that govern the accessory pathway of activation are the same rate parameters that govern the kinetics of deactivation. These linkages impact both structural and kinetic aspects of the measured inter-probe distance relaxations in the two activation measurements, the deactivation measurement and the \Cm-titration measurement. Recognizing these linkages enabled recovery of the model parameters from the measurements through a global analysis of the data because the observed experimental relaxation rates \emph{jointly depend upon} the linked model parameters. Independent empirical fitting of stopped flow experiments ignores the intensity information in the measurements and the inherent kinetic linkages in the system.  In the compartmental analysis of kinetic transitients---the most common type of analysis---intensity information is not utilized because (i) analytical expressions for intensity are complicated functions of the transition rates and initial conditions, (ii) initial conditions must be known or assumed, and (iii) the species-associated intensities, including intermediate species, must be known. The lack of constraint when data are independently fit and empirically fit to sums of exponentials makes drawing biological conclusions problematic.  In contrast, global analysis of a diverse set of experiments in terms of a single model facilitates strong conclusions because global analysis is self-constraining---recovered model parameters must account for the observed behavior in all experiments.

\subsection{Induced-fit opening}
Two mechanistic models of TnC-TnI allostery (Fig.~\ref{fig:models})---the population-shift-stabilization model and induced-fit model---are consistent with the small \Cm-induced change in isolated TnC and the large distance change observed upon saturation of TnC-TnI with \Cm~(Fig.~\ref{fig:time_resolved}). These models provide different interpretations with respect to structure, kinetics, and energetics of the transitions in the model of TnC-TnI allostery (Fig.~\ref{fig:network_models}b). The population-shift-stabilization model interprets the isomerization of TnC as a structural opening event that, in the absence of TnI, is energetically unfavorable and statistically improbable.  TnI switching, through TnI-R association with the opening-induced hydrophobic pocket in TnC, stabilizes the otherwise unfavorable transition.  The induced-fit model interprets the TnC isomerization as a structurally subtle, but energetically favorable, transition that generates a \Cm-primed state with a structure that is closed. TnC opening is the TnI-mediated opening of permissive \Cm-primed TnC.  The two models have different kinetic signatures.  Of the two, only the induced-fit model allows the rate of TnI-induced activation to exceed the rate of \Cm-induced activation.  The mechanism is diagnosed by an observed rate of TnI-induced activation (305 \sec, Fig~\ref{fig:activation}b) that greatly exceeds the slow rate of \Cm-induced activation (19 \sec, Fig~\ref{fig:activation}a).

Theories of allosterism assume that ligand agonists \emph{directly} stabilize the active state of their receptor (reviewed in \cite{Luque:2002bv}). In the population-shift-stabilization model, \Cm~binding does not directly stabilize the receptor TnC.  Instead TnC is \emph{indirectly} stabilized by \Cm~in a process that relies on a free energy drop provided by TnI. In the induced-fit mechanism, \Cm~binding directly stabilizes the primed-closed conformation of TnC, which we find has a 300-times greater affinity for \Cm~than the closed conformation. Our finding that the cardiac TnC-TnI assembly switches as an induced-fit process questions the belief that energetics must correlate with structural change, \textit{i.e.} small energy drops produce little structural change; large energy drops produce appreciable movement. TnC opening, which involves an 8.3 \AA~inter-probe distance change, is, thermodynamically, marginally favorable ($\Delta G = -1.1$ kJ/mol, 15 \Celcius). In contrast, the \Cm-priming step, which produces a subtle 2.2 \AA~inter-probe distance change, is thermodynamically very favorable ($\Delta G = -6.9$ kJ/mol, 15 \Celcius). Mechanistically, induced-fit opening appears to involve the replacement of intra-TnC hydrophobic interactions with inter-protein TnC-TnI hydrophobic interactions as the TnI-R diffusively creeps deeper into the TnC hydrophobic pocket. TnC opening is coupled to, and limited  by, TnI isomerization.

\subsection{Free energy landscape}
\begin{figure}
\centering
$\begin{array}{l}
\multicolumn{1}{l}{\mbox{\bf (a)}} \\ 
\includegraphics[width=10cm]{\FIGS/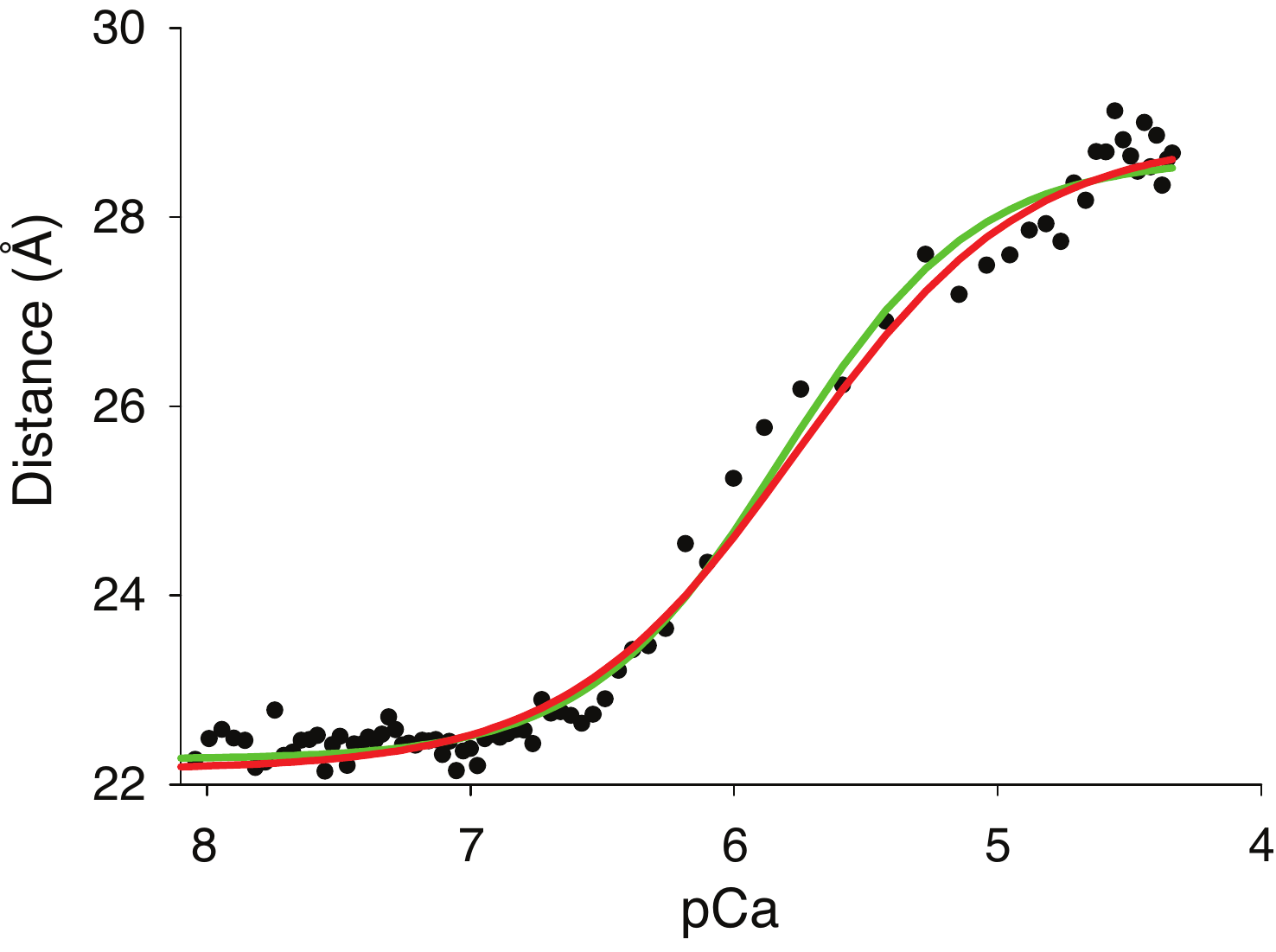} \\
\multicolumn{1}{l}{\mbox{\bf (b)}} \\ 
\includegraphics[width=10cm]{\FIGS/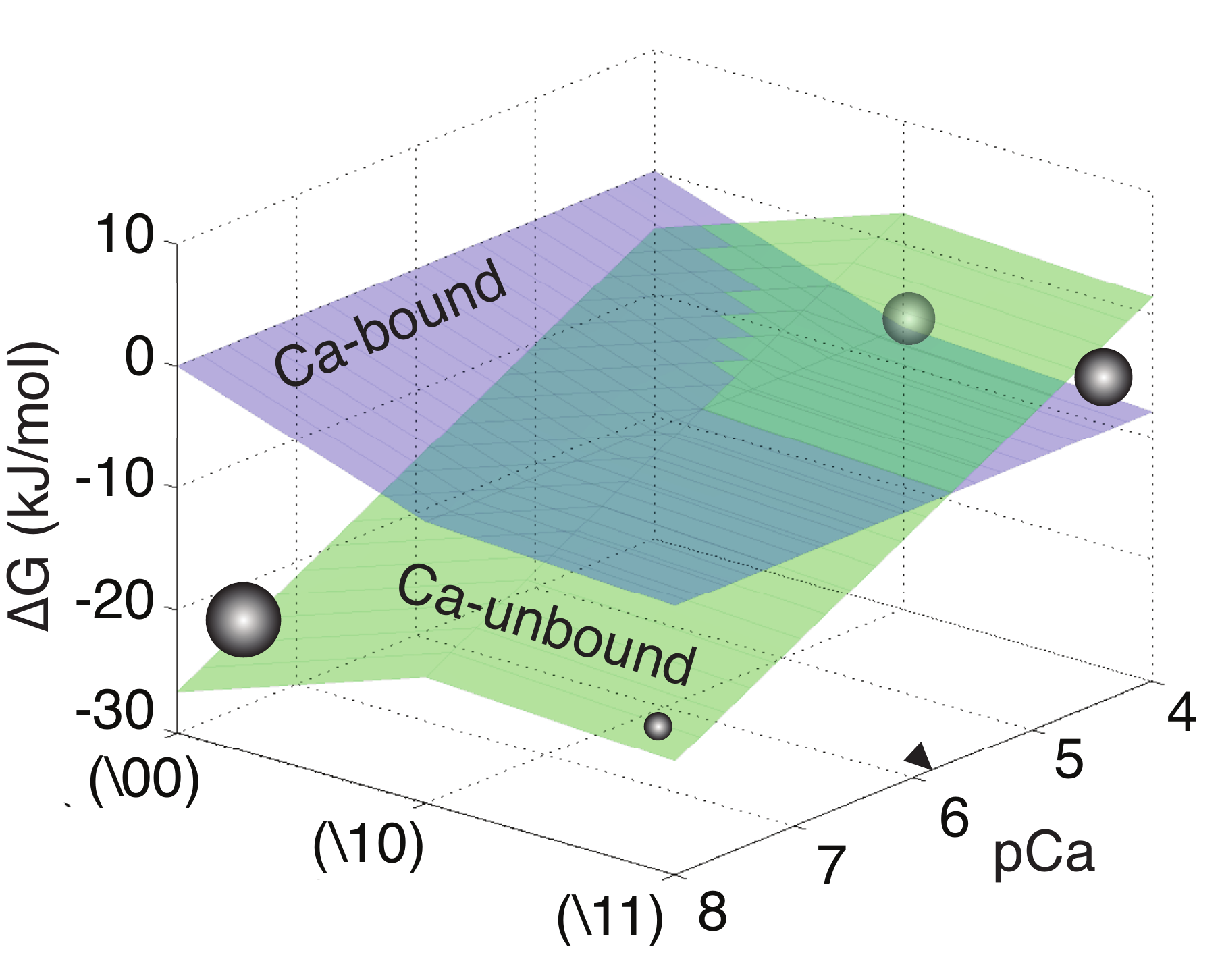} 
\end{array}$
\caption{
}
\label{fig:titration}
\end{figure}

\begin{figure}
\centering
\includegraphics[width=10cm]{\FIGS/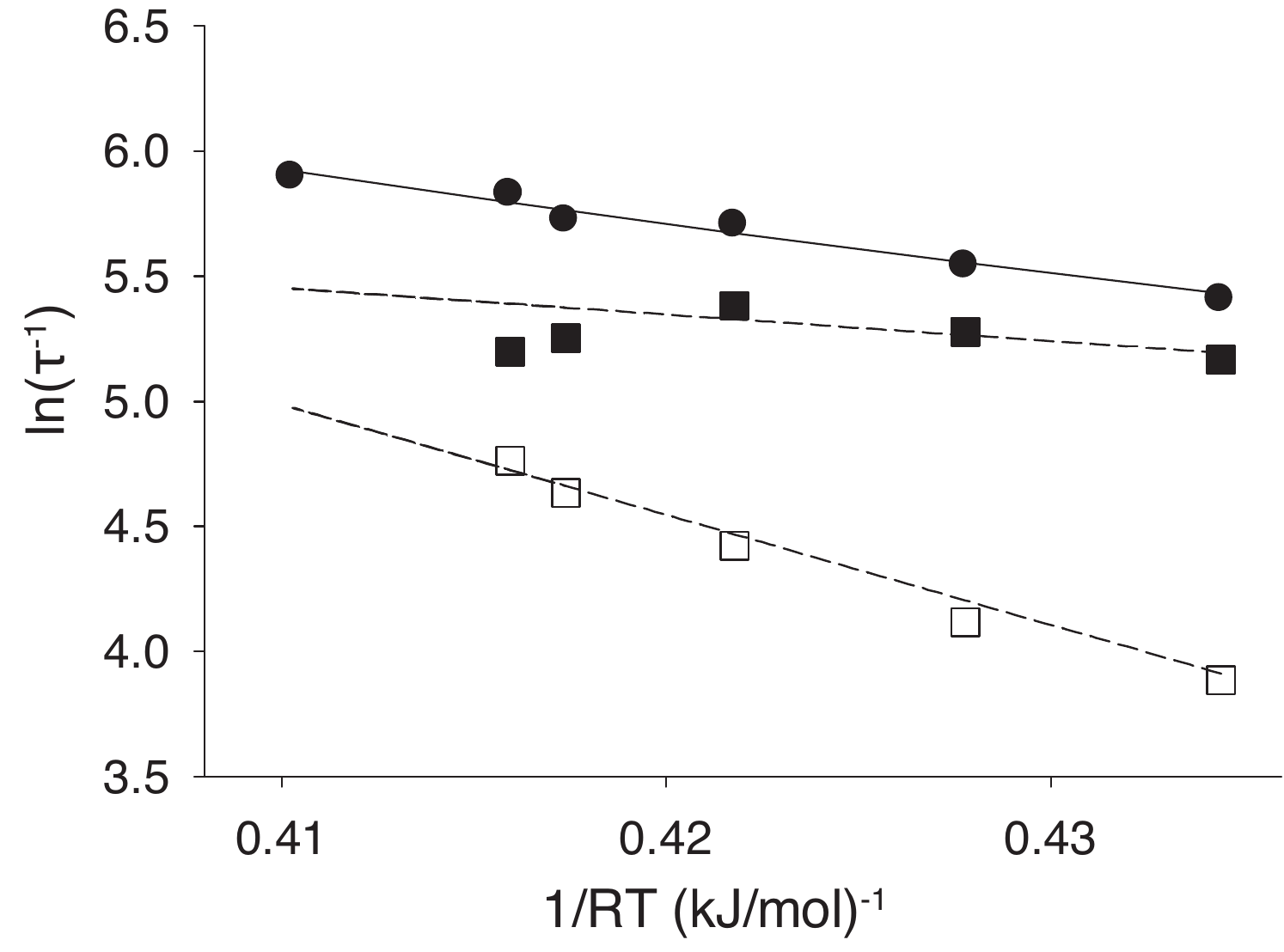}
\caption{
}
\label{fig:Arrhenius_plot}
\end{figure}

\begin{figure}
\centering
\includegraphics[width=10cm]{\FIGS/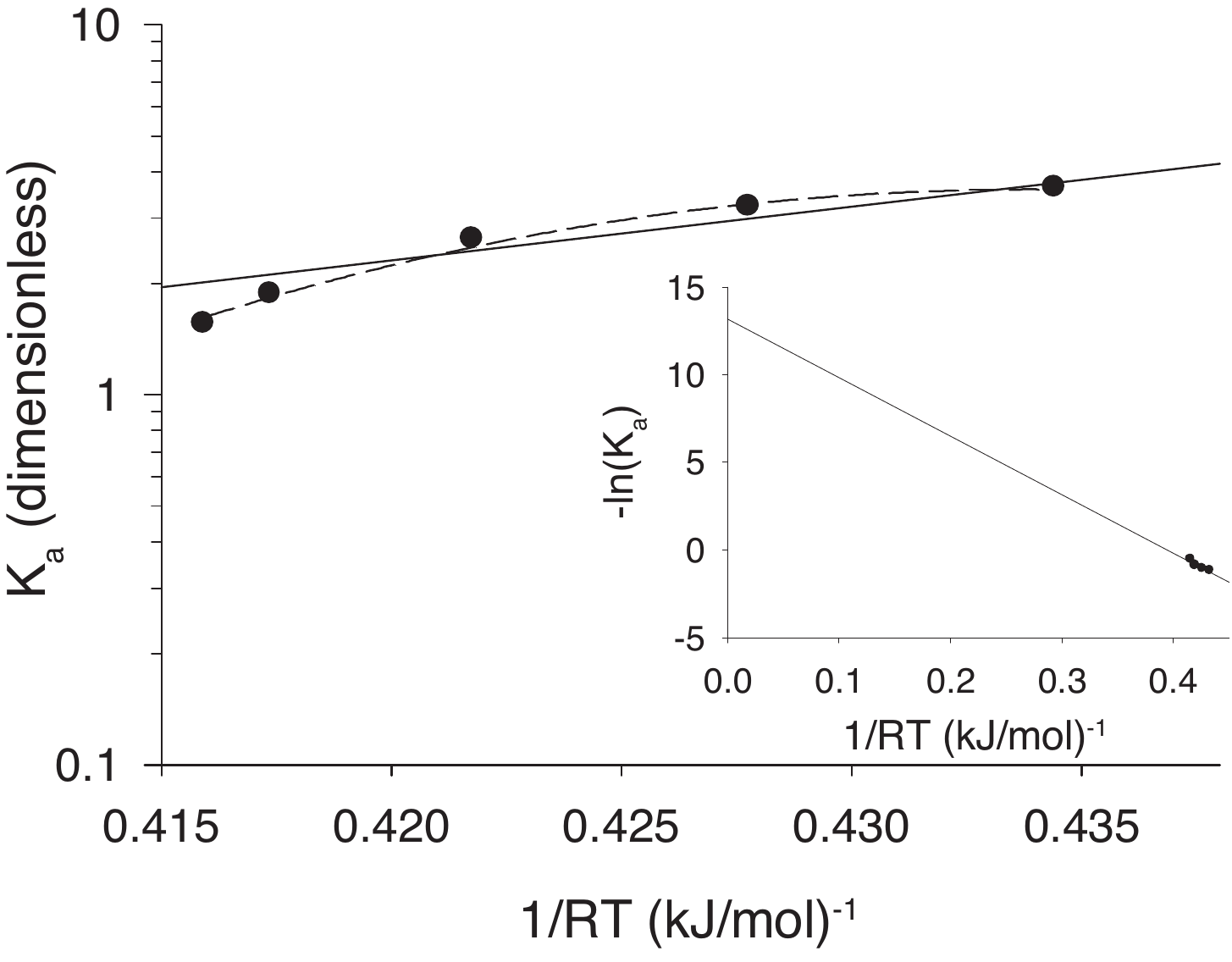}
\caption{
}
\label{fig:Vant_Hoff_plot}
\end{figure}

\begin{figure}
\begin{scheme}[H]
\centering
\includegraphics[height=3cm]{\FIGS/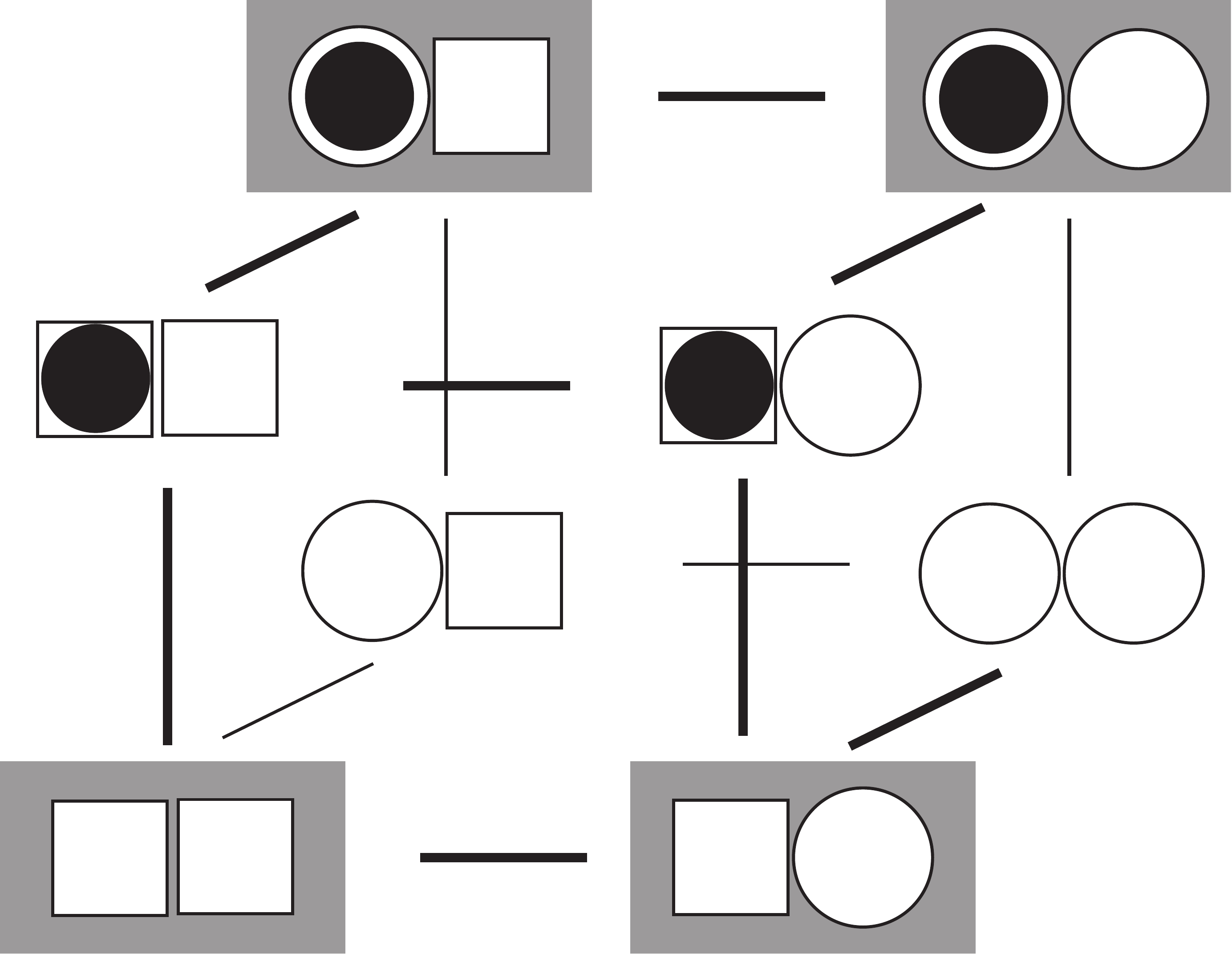}
\caption{ }
\label{scm:model}
\end{scheme}
\end{figure}

The \emph{mesoscopic free energy landscape} consists of the set of one dimensional free energy surfaces along the  reaction coordinates of all system-state transitions (edges in the Markov network, Fig.~\ref{fig:network_models}a). System-states (nodes in the Markov network) are regions of local stability (low energy wells) in the free energy landscape. Collectively, the system-states $S_i$ and the macroscopic free energies of the system states $G_{S_i}$ define the  \emph{macroscopic free energy landscape}, or \emph{free energy landscape} $\set{S_i, G_{S_i}}$ of the system. The free energy landscape of the TnC-TnI assembly (excluding the high energy states ($\bs01$)) is shown in Fig~\ref{fig:titration}b. The free energy landscape is of fundamental interest because it provides the weighting scheme needed to calculate the macroscopic measure of any observable property of the equilibrated system. Through the Boltzmann equation (Eq.~\ref{eq:Boltzmann}), the $G_{S_i}$ determine the equilibrium probability $P_{S_i}^\mathrm{eq}(\mu)$ that the system occupies system-state $S_i$ under an externally imposed \Cm~chemical potential $\mu$. The ensemble measurement of any observable $\mathcal{O}_{S_i}$ of the equilibrated system is  the inner-product of the observable with the  equilibrium probability distribution
\begin{equation}
\left< \mathcal{O}_{S_i} \right> (\mu)= \sum_{\set{S_i}} P_{S_i}^\mathrm{eq}(\mu) \mathcal{O}_{S_i}.
\label{eq:O}
\end{equation}
The $\mathcal{O}$-based $pCa_{50} = -\beta \mu'/\ln(10)$ is obtained  from the $\mu'$ where 
\begin{equation}
\widehat{ \left< \mathcal{O}_{S_i} \right> }(\mu') \equiv \frac{\left< \mathcal{O}_{S_i}  \right>(\mu')  - \left< \mathcal{O}_{S_i} \right> _{\min} }{ \left< \mathcal{O}_{S_i} \right> _{\max} -  \left< \mathcal{O}_{S_i} \right> _{\min} } = 1/2.
\label{eq:pD}
\end{equation}

The FRET distance $d_{S_i}$ (defined in Methods) in a \Cm ~titration experiment (\emph{c.f.} Fig.~\ref{fig:titration}a) is one such system observable $\mathcal{O}_{S_i}$; many others exist.  The distance-$pCa_{50}$ or $pd_{50}$(5.78) was obtained by solving Eq.~\ref{eq:pD} for the observable $d_{S_i}$. It is generally assumed that the sensitivity  to \Cm~of some physical observable of myofilament regulation (\emph{e.g.} force, probe fluorescence) is equal to the net affinity of the TnC regulatory site for \Cm. To test this assumption we examined the correspondence between the $pd_{50}$ and the net \Cm-binding-$pCa_{50}$ or $pK_{50}$. The \Cm-binding status $B_{S_i}$ of system-state $S_i$ is defined
\begin{equation}
B_{S_i} = \left\{ \begin{array}{cr}
1, & \mathrm{if}~s_0 = 1 \\
0,  &  s_0 = 0 
\end{array} \right..
\label{eq:Ca-bind}
\end{equation}
$\widehat{ \left< B_{S_i} \right> }(\mu)$ was calculated using Eqs.~\ref{eq:O},~\ref{eq:pD}, and~\ref{eq:Ca-bind}. Significantly, the traces of $\widehat{ \left< d_{S_i} \right> }(\mu)$ and $\widehat{ \left< B_{S_i} \right> }(\mu)$ superimpose, and the recovered $pd_{50}$ and $pK_{50}$ are identically 5.78 (data not shown). This supports the assumption that the sensitivity to \Cm~of a system observable is a measure of net \Cm~affinity for the regulatory site of TnC.

In addition to its quantitative value, the free energy landscape (Fig~\ref{fig:titration}b) provides a visual representation of how activation/deactivation occurs. The Boltzmann equation (Eq.~\ref{eq:Boltzmann}) dictates that the system will respond to any perterbation by preferentially populating the regions of lowest free energy. In the TnC-TnI assembly, as [\Cm] is raised, the $G_{S_i}$ of the \Cm-unbound system states $(0 \bs s_1 s_2)$ (green plane) rise above the \Cm-bound system states $(1 \bs s_1 s_2)$ (blue plane), which are stationary. Population density (represented as sphere radius in Fig~\ref{fig:titration}b) is greatest for those states with lowest free energy at a given [\Cm] (pCa). Activity is defined as a change in the probability that TnI, the reporting component of the assembly, is either active ($s_2 = 1$) or inactive ($s_2 = 0$). Visualized in terms of the free energy landscape, activation is rightward movement of density towards the TnI-active states $(0 \bs s_1 1)$ and $(1 \bs s_1 1)$. Deactivation is leftward movement of density away from the TnI-active states. At low [\Cm], the system is mostly deactivated in $(0 \bs 00)$ with very little partially activated species $(0 \bs 10)$. At high [\Cm] a 3:2 mixture of primed-closed $(1 \bs s_1 0)$ and open $(1 \bs s_1 1)$ system-states occur.

Although the micro-equilibrium constants $\mathcal{K}_1 = k_1 / k_{-1}=18.0$ and $\mathcal{K}_3 = k_3 / k_{-3}=0.052$ were determined inexactly, we can confirm that their values jointly satisfy the requirement that the work done on the free energy landscape by \Cm~binding not exceed the input of free energy to the assembly from \Cm~binding. From the \Cm~titration measurement, \Cm~binding provides $\Delta G_{in} = \ln(10)RT pCa_{50} = 31.9$ kJ/mol of free energy into the TnC-TnI assembly (15 \Celcius). This free energy is the maximum amount of energy available to change the free energy landscape of the assembly. Kinetic linkages governing the TnI isomerization, $(\bs 10) \leftrightarrow (\bs11)$, dictate that the bound and unbound surfaces have the same $\Delta G$ for the TnI isomerization, i.e. $G_{1\bs 11} - G_{1\bs 10} = G_{0\bs 11} - G_{0\bs 10}$. However, the \Cm-bound and \Cm-unbound surfaces of the free energy landscape have different $\Delta G$  for the TnC isomerization, $(\bs 00) \leftrightarrow (\bs10)$. Indeed, without some difference in $\Delta G$ between the \Cm-bound and \Cm-unbound surfaces, no \Cm-induced population shift would occur  for the protein states $(\bs s_1 s_2)$. From the estimated $\mathcal{K}_1$ and $\mathcal{K}_3$, the total relative change between the unbound and  bound free energy surfaces is $\Delta G_{out} = -RT \ln(\mathcal{K}_3/\mathcal{K}_1) = 14.0$ kJ/mol. The efficiency of coupling the free energy of ligand binding into a change in the free energy landscape, $\epsilon \equiv \Delta G_{out} / \Delta G_{in} = 0.44$. As required, $\epsilon \leq 1$.

\subsection{Thermo-kinetics of TnC-TnI opening}
From temperature-dependent stopped flow kinetics, we have determined the enthalpy $\Delta H$ (-33.4 kJ/mol), entropy $\Delta S$ (-0.110 kJ/mol/K), and heat capacity change $\Delta C_p$  (-7.6 kJ/mol/K) of the isolated opening transition in the activation of the cardiac TnC-TnI assembly (Table~\ref{tbl:landscape_params}).  The measured values are for the opening step in activation, which is distinct from other processes such as \Cm-binding, \Cm-priming, and overall TnC-TnI association.  Opening is promoted by a favorable decrease in enthalpy that is partially offset by a loss of entropy. Like cardiac TnC, calmodulin is a member of the super-family of bi-lobed \Cm~binding proteins with two EF-hand domains per lobe. The entropy change for the binding of a peptide of myosin light-chain kinase to calmodulin has been studied by deuterium NMR relaxation~\cite{Lee:2000qf}, where side chain order parameters provide an upper limit on $\Delta S$ (-0.4752 kJ/mol/K). Our measured $\Delta S$ (-0.110 kJ/mol/K) for TnI-R-induced TnC opening follows this trend of entropy loss upon complexation. Our measured $\Delta H$ and $\Delta S$ are opposite those observed for tension generating step in psoas muscle, which is endothermic and disordering~\cite{Wang:2001ei, Davis:2007lr}. Large negative $\Delta C_p$ have been consistently observed for ligand binding-induced isomerizations of proteins. Examples include $\Delta C_p =-3.5$ kJ/mol/K for DNA binding by the transcription factor Oct-1~\cite{Lundback:2000jh} and $\Delta C_p =-3.1$ kJ/mol/K for NAD$^+$ binding to glyceraldehyde-3-phosphate dehydrogenase~\cite{Niekamp:1977pc}. Large negative $C_p$ indicate  sequence-specific recognition~\cite{Prabhu:2005xr}. In most cases, $\Delta C_p$ are measured as average quantities for coupled equilibria involving combinations of bound/unbound ligand and active/inactive protein. In this context, $\Delta C_p$ is properly regarded as an apparent heat capacity change$\Delta C_p^{\mathrm{app}}$ with contributions from the intrinsic $\Delta C_p$ between the inactive and active protein states and the ligand-induced redistribution of the macrostate population~\cite{Eftink:1983wy}. Our measurements have resolved the true $\Delta C_p$ for the opening transition of the TnC-TnI assembly.

The measured $\Delta H$, $\Delta S$, and $\Delta C_p$ can be used to recover the distribution of enthalpy (energy) in  the open and primed-closed macrostates. The enthalpy distribution arises from  structural fluctuations---inter-converting microstates---within each macrostate.  Each macrostate consists of a set of $N$ microstates $m_1, m_2, \ldots, m_N$, and each microstate has an enthalpy $H(m_i)$. We define a set of discrete enthalpy intervals $H_1, H_2, \ldots, H_{\infty}$ of constant width $\Delta$. We cluster the microstate enthalpies $H(m_i)$ into the enthalpy levels $H_i$; each level contains some number $g_i$ microstates. The probability of finding the macrostate in the enthalpy interval $H_i$ is $p(H_i) = g_i /N$. The continuous probability distribution of enthalpy is obtained in the limit, $p(H) \mathrm{d}H = \lim_{\Delta \to 0}p(H_i)$. Evidently, $p(H)$ is the probability that a macrostate has enthalpy in the interval $[H,H + dH)$ \emph{given that $H$ is available} ($\sim e^{-\beta H}$, a decreasing function in $H$) times the probability that $H$ is available ($\Omega(H)$, an increasing function in $H$). $p(H)$ is therefore a peaked function. The enthalpy distribution $p(H)$ is parameterized by its moments 
\[
\left< {H}^j \right> = \int {H}^j p(H) \mathrm{d}H, \; j = 1,2, \ldots.
\]
Ignoring moments in $p(H)$ greater than two,  $p(H)$ is Gaussian 
\begin{equation}
p(H) \approx \frac{1}{ \sigma \sqrt{2 \pi}}\exp{\left[ - \frac{ \left( H- \left< H \right> \right)^2}{2 \sigma^2} \right]}
\label{eq:normdist}
\end{equation}
with mean $\left< H \right> = H^{\mathrm{obs}}$ and variance related to the enthalpy fluctuations $\delta H$ by $\sigma^2 = \left<  \delta H^2 \right> = \left< H^2 \right> - \left< H \right>^2 $. The variance is related to the heat capacity through
\begin{equation}
\sigma^2 =  R T^2 C_p.
\label{eq:var}
\end{equation}
By introducing the subscript $(k)$ into $p(H)$, $\left< H \right>$, $\sigma$, etc., we emphasize that the quantities are properties of the macrostate $(k)$---they are mesoscopic quantities---not macroscopic quantities of the system as a whole. The enthalpy distribution $p_{(k)}(H)$ of macrostate $(k)$ is thus parametrized by $H_{(k)} = \left< H \right>_{(k)}$ and ${C_p}_{(k)}$.

We now show how the measured $\Delta S$ for opening is used to resolve the heat capacities, hence the width of the enthalpy distributions, of the open and primed-closed macrostates. The macrostate, a mesoscopic entity, is a sufficiently large collection of microstates that thermodynamic quantities can be appropriately defined for it by statistical thermodynamics~\cite{Chandler_1987a}. As with macroscopic entropy, the mesoscopic entropy $S_{(k)}$ of macrostate $(k)$ is a functional of the probability distribution of enthalpy,
\[
S_{(k)} = -R \int_{- \infty}^{\infty} p_{(k)}(H) \ln p_{(k)}(H) \mathrm{d}H.
\]
For Gaussian distributed $p_{(k)}(H)$ (Eq.~\ref{eq:normdist}) with standard deviation $\sigma_{(k)}$, we obtain~\cite{Shannon:1948},
\begin{equation}
S_{(k)} = R \ln \sqrt{2 \pi e} \sigma_{(k)}.
\label{eq:gaus_entropy}
\end{equation}
Quite remarkably, $\sigma_{(k)}$ is completely determined by $S_{(k)}$. From Eqs.~\ref{eq:var}~and~\ref{eq:gaus_entropy}, the entropy change for opening, $ \Delta S = S_{(s_0 \bs 11)} - S_{(s_0 \bs 10)} $, is a simple expression of the heat capacities of the open and primed-closed macrostates,
\begin{equation}
 \Delta S = k_B \ln \sqrt{ \frac{{C_p}_{(s_0 \bs 11)}}{{C_p}_{(s_0 \bs 10)}} }.
\label{eq:S_Cp}
\end{equation}
 The specific heat capacities of the open and primed-closed states are 
 \begin{subequations}
\begin{equation}
 {C_p}_{(s_0 \bs 11)} = {C_p}_{(s_0 \bs 10)}  + \Delta C_p, \\
\end{equation}
\begin{equation}
{C_p}_{(s_0 \bs 10)} = \frac{\Delta C_p}{e^{2\Delta S / R} -1}.
\label{eq:Cp1}
\end{equation}
\end{subequations}
Evaluating, we obtain ${C_p}_{(s_0 \bs 10)} = 7.6$~kJ/mol and ${C_p}_{(s_0 \bs 11)} \sim 10^{-10}$~kJ/mol. The extremely low value of ${C_p}_{(s_0 \bs 11)}$ is due to the highly negative $\Delta S$, causing $e^{2\Delta S /R} \to 0$. The enthalpy distribution of the primed-closed state $p_{(s_0 \bs 10)}(H)$ is broad with $\sigma_{(s_0 \bs 10)} = 72$~kJ/mol ($T$ = 15 C). Compared to $p_{(s_0 \bs 10)}(H)$, the enthalpy distribution of the open state $p_{(s_0 \bs 11)}(H)$ is almost delta-correlated $\sigma_{(s_0 \bs 10)} \simeq 0$~kJ/mol, and shifted to lower enthalpy (by -33.4 kJ/mol).

We have obtained, additionally, the enthalpy of activation for opening $\delta^{\ddagger}$ (10.6 kJ/mol) and the entropy-adjusted, or effective, barrier crossing attempt frequency for opening $\nu_{\mathrm{adj}} \equiv \nu \exp \left( \sigma^{\ddagger} / R\right) $ ($1.8 \times 10^4$ \sec), where $\sigma^{\ddagger}$ is the entropy of activation. $\nu_{\mathrm{adj}}$ is very slow, and the rate of TnC opening is limited by a large loss of entropy $\sigma^{\ddagger}$ needed to reach the transition state. We conclude that TnC-TnI opening involves a large transient loss ($\sigma^{\ddagger} $) of configurational freedom and a net loss ($\Delta S$) of configurational freedom.

Finally, we note that the third law of thermodynamics demands that $C_p$ be non-negative~\cite{Chandler_1987a}. From Eq.~\ref{eq:Cp1} we obtain the constraint
\begin{equation}
\frac{\Delta C_p}{e^{2\Delta S / R} -1} \geq 0,
\label{eq:constraint}
\end{equation}
which is satisfied by our results. This provides a useful check for experimental results.

\section{Conclusions}
Our results indicate the presence of dynamic exchange on two levels: dynamic exchange among macrostates of the cardiac TnC-TnI regulatory complex (the closed, primed-closed, and open macrostates) and  dynamic exchange among microstates within each macrostate. A statistical description is necessary to explain the macroscopic (physiologic) behavior of cardiac regulation because activation and deactivation are never complete. The chance that the system occupies an active or inactive macrostate is only more or less probable. One thus speaks of the cardiac regulatory cycle in terms of shifting macrostate populations determined by a macroscopic free energy landscape that is modulated by the cytosolic concentration of \Cm.

\textbf{Acknowledgements} This work was supported by by the NIH (HL052508, HL80186), the European Union (MSCF-CT-2004-013119) and the American Heart Association (0330170N). Molecular structural images were rendered using VMD~\cite{Humphrey:1996lr}.




\section{Appendix A}
Here we derive expressions that allow an experimental thermo-kinetic parametrization of the energy landscape of a bi-meta-stable system that reversibly and stochastically jumps between two states, $A$ and $B$~(supplemental Fig. 1), according to the kinetic scheme
\begin{equation}
A \rates{k_{1}}{k_{-1}} B.
\label{eq:rxn}
\end{equation}
The reversible transition is of interest not only in single step reactions but also as a component step of multi-step reactions in coupled allosteric transitions. An Arrhenius analysis of temperature-dependent kinetic data, where the logarithm of the observed reaction rate is plotted against inverse temperature, is typically limited to the analysis of an irreversible reaction
\begin{equation}
A \rate{k_{1}} B\;  or \; A \revrate{k_{-1}} B
\label{eq:simp_rxn}
\end{equation}
either because, $k_{1} \gg k_{-1}$, or, $k_{1} \ll k_{-1}$, or because separate $k_{1}$ and $k_{-1}$ can be resolved through external information~\citep{Davis:2007lr}.  We wish to perform an Arrhenius analysis of the reversible first order process in Eq.~\ref{eq:rxn}.  A general expression is derived for the dependence of the observed relaxation rate, $\tau^{-1} = k_1 + k_{-1}$, on inverse temperature. The expression applies to both reversible and irreversible reactions. 

We assume that both forward and reverse isomerizations in Eq.~\ref{eq:rxn} are governed by the van't Hoff-Arrhenius law
\begin{equation}
k_{j \from i}  = \nu_{j \from i} \exp ( -\beta  \Delta G_{j \from i}^{\ddag} )
\label{eq:arrhenius}
\end{equation}
with effective barrier crossing attempt rate $\nu_{j \from i}$ from Kramers' reaction rate theory \citep{Kramers_1940a, Langer_1968a, Hanggi_1990a},  barrier crossing free energy $\Delta G_{j \from i}^{\ddag}$, and $\beta = 1/R T$. Barrier crossing free energies are defined by
\begin{equation}
\Delta G_{j \from i}^{\ddag} = \left\{ \begin{array}{cr}
\gamma^{\ddag} + \Delta G_{j \from i}, & \mathrm{if}~G_{i} < G_{j} \\
\gamma^{\ddag},  &  G_{i} \geq G_{j} 
\end{array} \right.
\label{eq:G_dagger}
\end{equation}
where $\gamma^{\ddagger}$ is the nominal Gibbs free energy of activation. Inserting Eq.~\ref{eq:G_dagger} into Eq.~\ref{eq:arrhenius}, the forward and reverse kinetic rates are given by
\begin{equation}
\begin{array}{cr}
k_1 = \nu_1 \exp ( -\beta \gamma^{\ddag} ) \\
k_{-1} = \nu_{-1} \exp [-\beta (\gamma^{\ddag} - \Delta G )]
\end{array}
\label{eq:rates}
\end{equation}
with forward  and backward barrier crossing attempt frequencies $\nu_{1}$ and $\nu_{-1}$. $\Delta G$, $\Delta G_{j \from i}^{\ddag}$, $\gamma^{\ddagger}$, $k_1$, and $k_{-1}$ are shown in supplemental Fig. 1. From detailed balance, $k_{-1}/k_{1} = \exp(\beta \Delta G)$, and we find that the forward and backward barrier crossing attempt frequencies are identical, $\nu_{1} = \nu_{-1} = \nu$. 

The observed relaxation rate $\tau^{-1}$ in any perturbation of a system governed by Eq.~\ref{eq:rxn} is the eigenvalue $\tau^{-1} = k_{1} + k_{-1}$. Substituting the expressions for $k_{1}$ and  $k_{-1}$ and rearranging, we obtain,
$\tau^{-1} = \nu \exp(-\beta \gamma^{\ddagger}) [1 + \exp(\beta \Delta G) ]$. The change in Gibbs free energy $\Delta G$ depends on the change in enthalpy $\Delta H$ and the change in entropy $\Delta S$ through, $\Delta G =  \Delta H - T  \Delta S$. Similarly, the change in the nominal activation free energy $\gamma^{\ddagger}$ depends on the nominal change in the enthalpy of activation $\delta^{\ddagger}$ and the nominal change in the entropy of activation $\sigma^{\ddagger}$ through, $\gamma^{\ddagger} = \delta^{\ddagger} - T \sigma^{\ddagger}$. To simplify the derivation we ignore any temperature dependence in $\Delta H$, $\Delta S$, $\delta^{\ddagger}$, and $\sigma^{\ddagger}$ over the limited experimental range of temperatures. Making the substitution, $\beta \gamma^{\ddagger} = \beta \delta^{\ddagger} - \beta T \sigma^{\ddagger} =  \beta \delta^{\ddagger} + \sigma^{\ddagger}$/R, we obtain
\begin{equation}
\tau^{-1} = \nu \exp \left( \sigma^{\ddagger} /R\right) \exp(-\beta \delta^{\ddagger}) [1 + \exp(\beta \Delta G) ],
\label{eq:master}
\end{equation}
where $\nu_{\mathrm{adj}} = \nu \exp \left(  \sigma^{\ddagger}/R \right)$ is the forward entropy-adjusted (effective) barrier crossing attempt frequency.
Taking the logarithm and defining $K_d = \exp(\beta \Delta G) = k_{-1}/k_{1} = x$, we obtain the equation for the curve of a reversible reaction in an Arrhenius plot
\[
\ln(\tau^{-1}) = \ln(\nu) + \sigma^{\ddagger} /R -\beta\delta^{\ddagger} + \ln(1+x).
\]
Recalling, $\partial \ln(1 + x) = (1 + x)^{-1} \partial (1 + x)$, and observing, $\lpd{x}{\beta} = x \Delta H$, we obtain the desired expression for the slope in an Arrhenius plot from the relaxation rate of a reversible reaction
\begin{equation}
\pd{\ln(\tau^{-1})}{\beta} =  - \delta^{\ddagger} + \frac{k_{-1}}{k_{1} + k_{-1}}\Delta H.
\label{eq:slope}
\end{equation}
In the forward irreversable limit $(k_{-1} \ll k_{1})$, we obtain the classical result, $\lpd{\ln(\lambda)}{\beta} = - \delta^{\ddagger}$. In the backward irreversable limit $(k_{-1} \gg k_{1})$, $\lpd{\ln(\lambda)}{\beta} = - \delta^{\ddagger} + \Delta H$. In a fully reversable reaction, $(k_{-1} = k_{1})$, $\lpd{\ln(\lambda)}{\beta} = - \delta^{\ddagger} + \Delta H/2$. In an Arrhenius analysis that ignores reversibility, activation enthalpy is underestimated for endothermic reactions ($\Delta H > 0$), and activation enthalpy is overestimated  for exothermic reactions ($\Delta H < 0$).

\section{Appendix B}
The equilibrium association constants $K_0$, $K_1$, and $K_2$ for \Cm~binding to the system states $(\bs 00)$ , $(\bs 10)$, and $(\bs 11)$ are defined based on two assumptions. \emph{(i)} The free energies of all ligand-unbound species $G_{0 \bs s_1 s_2}$ depend linearly on the chemical potential of [\Cm], $\mu=RT \ln[\mathrm{Ca}^{2+}]$, and the free energy of the protein complex $G^o_{0 \bs s_1 s_2}$: $G_{0 \bs s_1 s_2} =  G^o_{0 \bs s_1 s_2} + n \mu$ ($n = 1$). \emph{(ii)} The free energies of the ligand bound species $G_{1 \bs s_1 s_2}$ are fixed:   $ G_{1 \bs s_1 s_2} =  G^o_{1 \bs s_1 s_2}$ ($n = 0$). For \Cm~binding to some configuration of TnC-TnI $(\bs s_1 s_2)$, Eq.~\ref{eq:probs} gives
\[
\left.
\begin{array}{lcl}
\lfrac{P^{\mathrm{eq}}_{1 \bs s_1 s_2}}{P^{\mathrm{eq}}_{0 \bs s_1 s_2}}  & =  & e^{-\beta \left(G_{1 \bs s_1 s_2} -  G_{0 \bs s_1 s_2}\right)} \\
  & = &  e^{-\beta \left( G^o_{1 \bs s_1 s_2} -  G^o_{0 \bs s_1 s_2}  - \mu \right)} \\
  & =  &  e^{\beta \mu}e^{-\beta \left(  G^o_{1 \bs s_1 s_2} -  G^o_{0 \bs s_1 s_2} \right)} \\
  &  = &    [\mathrm{Ca}^{2+}]e^{-\beta \ \Delta G^0_{\bs s_1 s_2}  }
\end{array}
\right. .
\]
where $ \Delta G^o_{\bs s_1 s_2} = G^o_{1 \bs s_1 s_2} - G^o_{0 \bs s_1 s_2}$. The equilibrium association constant $K_0$ for \Cm~binding to $(\bs 00)$ is
\[
K_0 = W_+/W_{-} = e^{-\beta  \Delta G^0_{\bs 00}  },
\]
and
\[
K_0[\mathrm{Ca}^{2+}] = [\mathrm{Ca}^{2+}] W_+ / W_{-} = \lfrac{P^{\mathrm{eq}}_{1 \bs 00}}{P^{\mathrm{eq}}_{0 \bs 00}}.
\]
Thus, $K_0[\mathrm{Ca}^{2+}] $ is the pseudo-equilibrium constant that governs \Cm~binding to fully deactivated TnC-TnI (i.e. the $(0 \bs 00) \to (1 \bs 00)$ transition). 
$K_1[\mathrm{Ca}^{2+}] $ and $K_2[\mathrm{Ca}^{2+}] $ are the pseudo-equilibrium constants that govern \Cm~binding to partially activated $(0 \bs 10)$ and fully activated $(0 \bs 11)$ TnC-TnI. $K_1$ and $K_2$ are related to $K_0$ by macroscopic detailed balance (Eq.~\ref{eq:MDB}): $K_0 W_1 W_{-3} = K_1 W_3 W_{-1}$ and $K_1 W_2 W_{-2}= K_2W_2 W_{-2}$, which simplifies to $K_1 = K_2$.


\newpage
\clearpage
\section{Tables}

\begin{table}[c]
\caption{Microscopic parametrization of the \Cm-signaling cycle. Parameter values are from a global fit of activation and deactivation stopped flow FRET data (c.f. Figs.~\ref{fig:activation} and~\ref{fig:deactivation}) to the model of the signaling cycle (Fig~\ref{fig:network_models}, Eqs.~\ref{eq:activation} and~\ref{eq:deactivation}) (15 \Celcius). Reported values are mean $\pm$ precision.  Precision is expressed as $\set{le, ue}$, where $le$ and $ue$ are the respective lower and upper 68\% (1 $\sigma$) confidence estimates of the parameter, obtained from a 1-D adiabatic grid search. $^a$ fixed (see text). $^b$ poorly resolved, confidence estimates unavailable. $^c$ precision not determined because value depends on assumed $k_{-1}$. $^d$ fixed to value measured from time-resolved FRET ($\overline{d_{10}} \pm \sigma$). 
}
\begin{center}
\begin{tabular}{|l c |}
\hline
Parameter & global fit \\
\hline
$k_{1}$ (\sec) & $18.0 \pm \set{14.4, 21.9}$  \\
$k_{-1}$ & $1.0^a$ \\
$k_{2}$ & $181 \pm \set{145, 222}$ \\
$k_{-2}$ & $117 \pm \set{93, 148}$ \\
$k_{3}$ & $1618^b$ \\
$k_{-3}$ & $31,000^a$ \\
$d_{00}$ (\AA) & $21.4^c$  \\
$d_{10}$ & $23.9 \pm 2.6^d$ \\
$d_{11}$ & $32.2 \pm \set{31.3, 34.5}$ \\
\hline
\end{tabular}

\end{center}
\label{tbl:params}
\end{table}%

\newpage
\clearpage
\begin{table}[c]
\caption{Thermo-kinetic parametrization of the opening/closing transition of cardiac TnC-TnI. For opening: $\Delta H$, net enthalpy change; $\Delta C_p$, heat capacity change; $\Delta S$, net entropy change; $\delta^{\ddagger}$, enthalpy of activation; $\nu_{\mathrm{adj}} =\nu \exp \left( \sigma^{\ddagger}/R \right)$, entropy-adjusted barrier crossing attempt frequency ($\nu$, barrier crossing attempt frequency; $\sigma^{\ddagger}$, entropy of activation.)  For closing:  $\delta_{\mathrm{R}}^{\ddagger}$, enthalpy of activation; and $\nu_{\mathrm{adj, R}}$ entropy-adjusted barrier crossing attempt frequency.
}
\begin{center}
\begin{tabular}{|l c|}
\hline
Parameter & Value \\
\hline
$\Delta H$ (kJ/mol) & -33.4 \\
$\delta^{\ddagger}$ & 10.6\\
$\delta_{\mathrm{R}}^{\ddagger}$ & 44.0\\
$\Delta S$ (kJ/mol/K)&  -0.110 \\
$\Delta Cp$ & -7.6\\
$\nu_{\mathrm{adj}}$ (\sec) & $1.8\times 10^{4}$\\
 $\nu_{\mathrm{adj, R}}$ & $1.0 \times 10^{10}$\\
\hline
\end{tabular}
\end{center}
\label{tbl:landscape_params}
\end{table}%


\newpage
\clearpage
\section*{Figure Legends}

\subsubsection*{Figure~\ref{fig:MD_sims}.}
FRET assay for the structural kinetics of the CRS. The engineered FRET donor Trp12 (D), the FRET acceptor Cys51-labeled AEDANS (A), and the N-terminal portion of TnC (ribbon) are shown. The efficiency of photon transfer between the emission dipole of Trp12 and the absorption dipole of AEDANS (shown) depends on the inter-probe distance (R), which changes upon hinge-like movement of helix B of TnC ($\alpha_{\mathrm{B}}$) during activation. (a) \Cm-bound TnC in the absence of TnI (modified from PDB ID code 1AP4). (b) The \Cm-bound TnC with bound TnI-R (brown helix) (modified from PDB ID code 1J1E). 

\subsubsection*{Figure~\ref{fig:network_models}.}
The TnC-TnI assembly represented as a Markov network (see text). (a) The parent network with system-states $(s_0 \bs s_1 s_2)$ as nodes and bi-directional transitions as edges. (b) Reduced models of activation and deactivation showing transition rates $k_i, i = 1,2,3$, \Cm~affinity constants $K_j, j=0,1,2$, kinetic linkages between activation and deactiation, and degeneracy in $k_2$ and $k_{-2}$. Dominant (red) are minor (blue) pathways and common portions (green) are shown. Double arrows ($\Leftrightarrow$) indicates transitions with the reversible binding of \Cm.  For protein isomerizations, single arrows indicate the direction of net probability flux.

\subsubsection*{Figure~\ref{fig:time_resolved}.}
Recovered FRET distance distribution.  Background-corrected time-resolved donor-only and donor-acceptor decays were fit to a Gaussian distributed population of inter-probe distances with mean (bar) and standard deviation (shown as an error bar). Samples: starting \Cm-free samples of TnC-TnI complex or isolated TnC (bold) with addition of saturating \Cm~or sequential addition of saturating \Cm~then TnI.

\subsubsection*{Figure~\ref{fig:models}.}
Mechanistic models of TnC-TnI activation consistent with the distance changes in Fig.~\ref{fig:time_resolved}.  (a) Population-shift-stabilization model: TnC first undergoes an energetically unfavorable opening event (1), which is subsequently stabilized by TnI-R binding (2). (b) Induced-fit model: \Cm~binding produces an energetically favorable \Cm-primed species that is not open (1). In the presence of TnI, the \Cm-primed complex undergoes concerted TnC opening and TnI-R binding (2). The kinetic signatures of each model are shown, with relative arrow length indicating relative transition probabilites.

\subsubsection*{Figure~\ref{fig:activation}.}
Activation kinetics. FRET optical distance versus time, 15 C. (a) \Cm-induced activation: change following rapid mixing of preformed binary TnC-TnI complex with \Cm~solution (top dotted trace) or with buffer (mock injection, bottom trace). Empirical double exponential fit (green):   $\tau_{1}^{-1}$ = 1554~\sec (48\% amplitude), $\tau_{2}^{-1}$ = 19~\sec (52\%). Global model-based fit (red) with derived parameters in Table~\ref{tbl:params}.  (b) TnI-induced activation: change following rapid mixing of TnI into \Cm-pre-saturated TnC. Empirical single exponential fit (green): $\tau^{-1}$ = 305~\sec. Global model-based fit (red) with derived parameters in Table~\ref{tbl:params}.

\subsubsection*{Figure~\ref{fig:deactivation}.}
Deactivation kinetics after \Cm~removal, 15 C. FRET optical distance change after rapid mixing of  \Cm-saturated TnC-TnI with buffer containing \Cm~chelator, EGTA. Empirical single exponential fit (green, obscured by red trace):   $\tau^{-1}$ = 125~\sec. Global model-based fit (red) with derived parameters in Table \ref{tbl:params}.  Inset: $\overline{\chi^2}$~(Eq.~ \ref{eq:chi_sqr}), the normalized adiabatic projection of the global $\chi^2$ hyper-surface on a model parameter  (labeled) axis.

\subsubsection*{Figure~\ref{fig:titration}.}
System response to a change in [\Cm]. (a) \Cm~titration, mean FRET distance \emph{vs.} $pCa$ (-log([\Cm])). Empirical fit to the Hill equation (green) ($n = 1.18$; $pCa_{50} = 5.82$). Global fit to the model (Eqs.~\ref{eq:marginal},~\ref{eq:R},~\ref{eq:Boltzmann}) (red) ($\beta \Delta \mu' = 22.9$ kJ/mol). (b) The experiment-resolved BFEL (see text, T = 15 \Celcius). \Cm-unbound system-states (green). \Cm-bound system-states (blue). Population density (spheres, radius represents magnitude) at the extremes of activation. $pCa_{50}$ (labeled, $\blacktriangleright$).

\subsubsection*{Figure~\ref{fig:Arrhenius_plot}.}
Arrhenius analysis of the induced fit opening step of activation. The rate of TnI-induced activation (circles), $\tau^{-1} = k_2 + k_{-2}$, at $T = \set{3.7, 10, 15, 16, 20}$ \Celcius~(c.f. Figure~\ref{fig:activation}b) were fit to Eq.~\ref{eq:Arrhenius_fit} (solid line) to recover the enthalpy of activation $\gamma^{\ddagger}$ and the entropy-adjusted barrier crossing attempt frequency $\nu_{\mathrm{adj}}$ for the opening transition (Table~\ref{tbl:landscape_params}).  Also shown (squares) are the $k_2(T)$ (filled) and $k_{-2}(T)$ (open) recovered from global analysis of relaxation data at  $T = \set{3.7, 10, 15, 16}$ \Celcius~(c.f. Figures~\ref{fig:activation} and~\ref{fig:deactivation}) and calculated $k_2$ and $k_{-2}$ (dashed lines) from the recovered $\nu_{\mathrm{adj}}$, $\nu_{\mathrm{adj, R}}^{\ddagger}$, $\gamma^{\ddagger}$, $\gamma_{\mathrm{R}}^{\ddagger}$.

\subsubsection*{Figure~\ref{fig:Vant_Hoff_plot}.}
Thermodynamic analysis of the induced fit opening step of activation. The temperature-dependent equilibrium constant, $K_a = k_{2}/k_{-2}$, (circles) from $k_2(T)$ and $k_{-2}(T)$ (Fig.~{\ref{fig:Arrhenius_plot}}) were fit using Eq~\ref{eq:Vant_Hoff_fit} with temperature-independent $\Delta H$ and $\Delta S$ (solid line) and temperature-dependent $\Delta H$ and $\Delta S$ (dashed line). Recovered  $\Delta H$, $\Delta S$, and $C_p$ are given in Table~\ref{tbl:landscape_params}. Inset: van't Hoff plot showing extrapolated results from fit to temperature-independent $\Delta H$ and $\Delta S$. The y-intercept and slope provide $-\Delta S/R$ and $\Delta H$, respectively.

\subsubsection*{Scheme~\ref{scm:model}.}
Coarse-grained model of an allosteric protein.

\newpage
\clearpage
\section{Figures and Schemes}
\newpage
\clearpage

\newpage
\clearpage

\newpage
\clearpage

\newpage
\clearpage

\newpage
\clearpage

\newpage
\clearpage

\clearpage

\newpage
\clearpage

\newpage
\clearpage

\newpage
\clearpage

\clearpage
\newpage
 \section*{Supplemental Online Data}
\clearpage


\begin{sfigure}
\centering
\includegraphics[width=10cm]{\FIGS/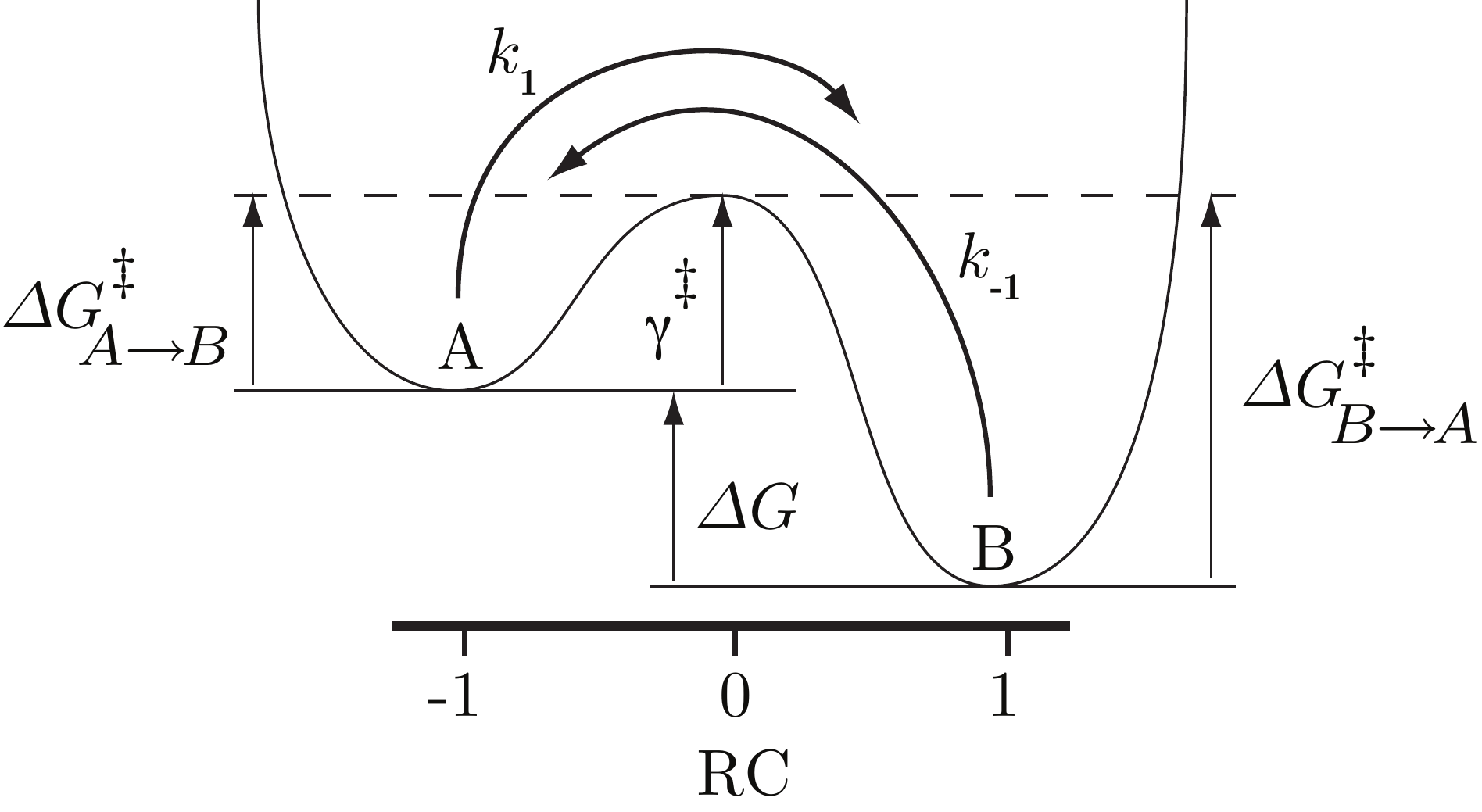}
\caption{Energy surface of a reversible noise-induced transition in a bi-metastable potential.
}
\label{fig:landscape}
\end{sfigure}

\begin{sfigure}
\centering
\includegraphics[scale=0.7]{\FIGS/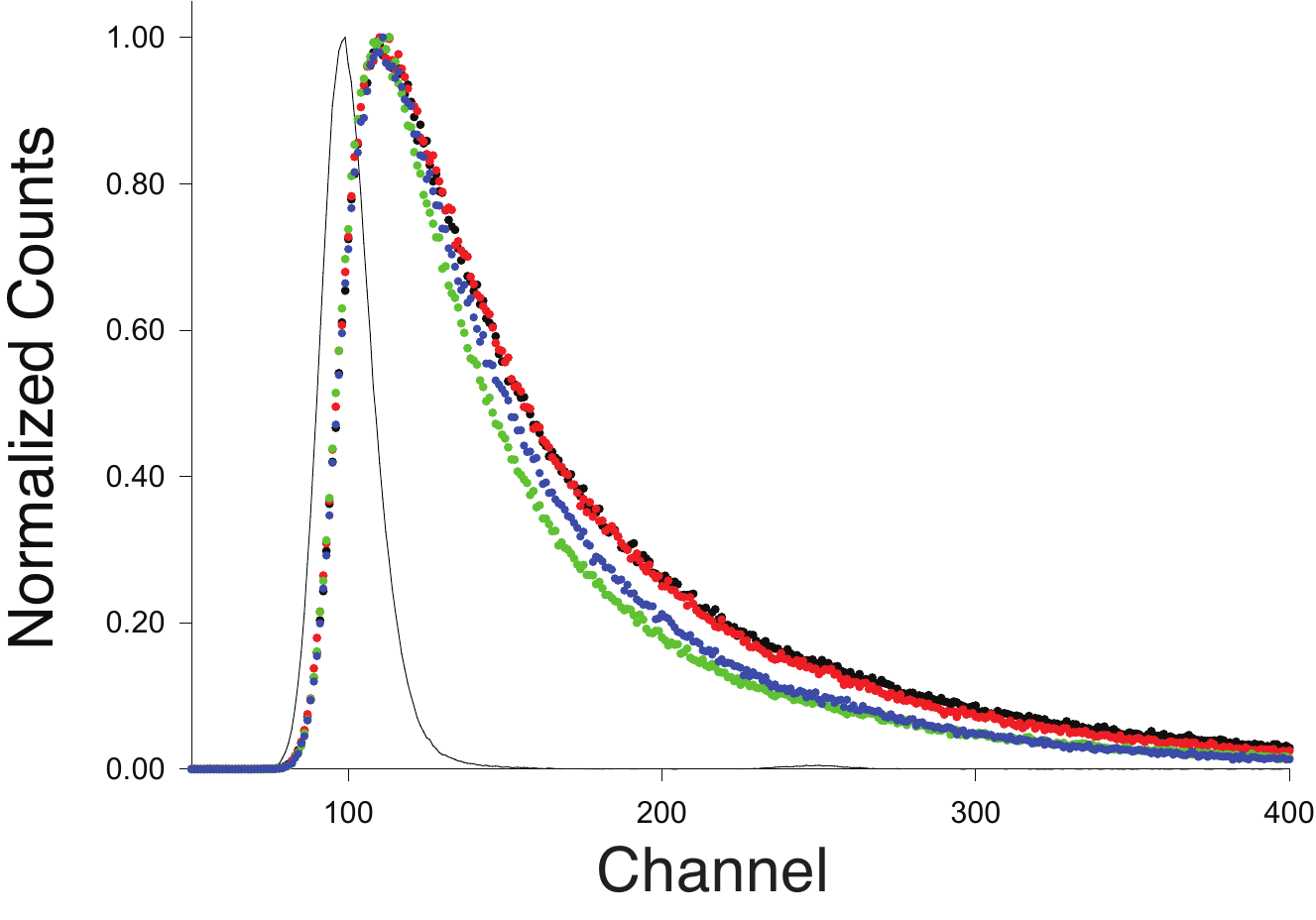}
\caption{
Time resolved decays of Trp12 in TnC(12W/C51$\pm$AEDANS) showing the effect of TnI addition (48.8 ps/channel). TnC in 160 \uM~\Cm: donor-only (black), donor-accepter (green). TnC in 160 \uM~\Cm~+ 2-fold excess of TnI: donor-only (red). Donor-acceptor (blue). Impulse response (black line).
}
\label{fig:decays}
\end{sfigure}

\end{document}